\documentclass[
reprint,
amsmath,amssymb,aps,prstper]{revtex4-2}

\usepackage{graphicx}
\usepackage{dcolumn}
\usepackage{bm}

\begin{document}

\title{Gender Equity in Physics Labs: A Review}

\author{Danny Doucette}
 \email{danny.doucette@gmail.com, he/him/his}
\affiliation{Department of Physics, St. Francis Xavier University, Antigonish, Nova Scotia, B2G 2W5, Canada} 
\author{Chandralekha Singh}
\affiliation{Department of Physics and Astronomy, University of Pittsburgh, Pittsburgh, Pennsylvania 15260, USA}

\date{\today}

\begin{abstract}

This review article provides an overview of research on the topic of gender equity in educational physics labs. As many institutions and instructors seek to evolve or transform physics lab learning, it is important that changes are made that improve equity for all students along multiple axes of identity, including gender. The studies highlighted in this review article describe the existence of complex gender-based differences, e.g., in opportunities to tinker with lab equipment, as well as differences in grades, conceptual understanding, and motivational outcomes across a broad range of lab curricula and contexts. The studies also illustrate and explore social interactions and structures that can impact students' experiences based on their gender identities. Although there has been less scholarship focused on proposals to reduce gender-based inequities in labs, this review article also provides an overview of some relevant proposals as well as associated research results. This overview of research on gender equity in physics labs helps to make clear that future scholarship on equity in physics labs should adopt gender frameworks that allow researchers to transcend binary gender identities and student deficit framing of research results. Likewise, a case is made that future research is needed on equity along other axes of identity, as well as research that accounts for intersectionality of different identities, in the physics lab context.

\end{abstract}

\maketitle

\section{Introduction}

This paper provides a review of research on gender equity in educational physics labs. Lab-work has long been a key component of physics learning in Western-style education systems~\cite{DeboerHistory}. However, despite decades of access, there continue to be challenges facing people who do not have a masculine identity when they enroll in physics lab courses. Our goal is to present two aspects of this story: first, to summarize research focused on identifying gender-based differences in physics labs, and second, to provide an overview of scholarship that addresses proposals to reduce gender inequities in physics labs.

Since this review summarizes prior research on gender equity in the physics lab context, we would like to start by explaining what we mean by equity in physics learning. Our conceptualization of equity in physics learning includes three pillars: equitable access and opportunity to learn physics, an equitable and inclusive learning environment, and equitable outcomes (e.g.,~\cite{CwikFramework}). Thus, by equity in physics learning in the lab context, we mean that all students should have equitable opportunities to learn and equitable access to resources. The learning environment needs to have appropriate support so that all students can engage in all aspects of learning and developing skills (including specific lab-related skills) in a meaningful and enjoyable manner. Likewise, the learning outcomes should be equitable. By equitable outcomes, we mean that students from all demographic groups (e.g., regardless of their gender identity or other identities) who have the pre-requisites to enroll in physics courses have comparable learning outcomes including the opportunity to develop lab-related skills. The learning outcomes, in this sense, could refer to grades, results from concept inventories, etc. as well as motivational outcomes such as self-efficacy and skill-development.  This conceptualization of equitable outcomes is consistent with Rodriguez et al.’s equity of parity model (i.e., students from all demographic groups have similar end-of-term outcomes) and is different from the equal gains in outcomes model (i.e., all students improve by the same amount so those who started ahead at the beginning of a course remain ahead at the end)~\cite{RodriguezEquityModels}.

Ensuring equitable outcomes based upon the equity of parity model requires an equitable and inclusive learning environment and intentional focus on ensuring that students from all demographic groups are provided opportunities to engage in and learn from all aspects of lab. We rely on Cochran's description of inclusion as being ``concerned with what being included in that organization or environment means''~\cite{CochranKitchen}, rather than the simple issue of whether all students are allowed to participate. Inclusiveness of learning environment is essential for it to be equitable but unless outcomes are equitable, the learning environment is not equitable. Equity of parity in outcomes typically requires modifications to instruction so that all students are given opportunities to engage in all aspects of physics labs in an equitable and inclusive learning environment. This type of environment can elevate the outcomes of groups (and ensure growth of all individuals in the group) regardless of whether they initially have had less relevant opportunities and had, e.g., lower-levels of prior preparation or self-efficacy.  When we refer to past studies that compare the outcomes for women and men, we are employing an equity of parity framework to deem the outcomes equitable. We note that several of the studies we describe outcomes by simply making comparisons between students from two gender groups. This is not aligned with our framework of equity, which requires analysis of issues of access and the learning environment, in addition to comparing performance on some metric of different demographic groups, e.g., women and men. Scholarship that consists of gender comparisons without an analysis of the learning environment and social dynamics is inadequate and counterproductive~\cite{GutierrezGapGazing}.

We focus in this paper on physics labs in a broad sense. This includes different educational levels, such as high school, college, `beyond the first year' of physics studies at university, and research labs that include an educational element (such as labs that train graduate students). We include studies with different lab configurations: separate or standalone lab courses, as well as integrated `studio physics'-style courses. The studies we cite below focus on a wide range of different curricula, including skills-based labs, conceptual inquiry-based labs, project-based labs, computational labs, and more~\cite{HolmesLandscape}. Physics lab learning is an essential component of physics education that includes instruction related to the epistemology and skills associated with experimental physics. Compared with lecture-based learning, lab-work offers students a different way of developing a sense of oneself as a physicist, and so can be important for students' physics identity development.

Regardless of the configuration or curricula, physics labs tend to be spaces in which students are expected to work collaboratively with their peers. As a consequence, physics labs are a place in which the culture of physics is learned, practiced, and deployed. By the culture of physics, we are referring to the constellation of stereotypes and ideas about what counts as physics and who counts as a physicist. This includes stereotype threat~\cite{SteeleWhistling,MariesAgreeing}, brilliance attributions~\cite{LeslieBrilliance}, and media portrayals of physicists~\cite{WeitekampBigBangTheory}. Physics differs from other disciplines because of the depth to which it is committed to masculinity~\cite{WertheimPythagoras,HasseGender}, especially in lab settings where collaboration and norms make gender-based biases and stereotypes especially prevalent and impactful~\cite{GonsalvesMasculinities}. Gender-based inequities are a significant issue in physics lab education, as they undermine our shared conviction that people from all demographic groups should have the opportunity to develop skills (including those specific to labs that have the potential to increase their self-efficacy related to science in general and experimental science in particular) as well as learn physics concepts in an equitable and inclusive learning environment. Lack of equity and inclusion in physics lab learning environments can deprive students, e.g., those from traditionally marginalized groups such as women, of the opportunities to develop those skills that could also have negative consequences on their academic outcomes as well as future career path. Thus, scholarship about the origins of gender-based inequities in physics labs, and especially scholarship focused on proposals to reduce and eliminate inequities, is necessary.

Our approach to gender and gender-focused research acknowledges gender as a fluid, non-binary, complex aspect of identity~\cite{TraxlerEnriching}. In the context of physics learning, Danielsson et al. posited gender as a practice~\cite{DanielssonLearning}: gender is something we \textit{do}~\cite{WestDoingGender}, in a context-dependent way, as well as a framing through which we interact with people and social systems around us. Within this perspective, we further acknowledge the prevalence of two particular genders (man and woman) and their associated structuring practices~\cite{ConnellMasculinities} (i.e., masculinity and femininity). Masculinity, then, is the pattern of ways that men seek to behave as men, and the social norms and ideas that emerge from that. Other gender identities exist, but these two are common enough to have motivated substantial scholarship. These two genders have been employed in a binary form by many scholars of past research. In this paper, we choose to respect the authorial decisions of the scholars we cite by using their terminology while describing their work. We hope that the reader can make it through the resulting mish-mash, e.g., of gender and sex being used interchangeably even though `'female' and `male' describe the sex of an individual rather than socially constructed gender, and use of gender binary even though gender is a fluid non-binary construct. We also use the phrase gender gap, which is employed by researchers to describe differences between the physics lab outcomes of women and men. We note that in all of the scholarship described below men outperformed women according to the relevant outcome in any case of a gender-based difference in that outcome.

Prior research suggests that a major underlying cause of most gender differences observed in physics labs is the masculine culture of physics~\cite{HallChillyClimate,SeymourTalking}. The masculine culture serves as the foundation for relationships between peers in physics settings, sets the tone for instructor-student interactions, and shapes norms and policies. The fact that masculinity dominates physics culture is true for all of the national contexts described in the studies we review, although the mechanisms for this domination may vary somewhat. Likewise, the ways that physics culture show up vary for different types of institution, and so it is important not to over-generalize the results from one study~\cite{KanimDemographics}. Within our framework, however, we reject claims about gender differences in physics labs that are rooted in biological differences and do not focus on how the interplay of social dynamics and gender affects student outcomes.

This review focuses on only one dimension of identity, i.e., gender, rather than treating equity along other student identities in a multi-dimensional space as well as intersectionality. This is necessitated by the shortage of scholarship focused on equity in physics labs along other dimensions such as race, ethnicity, disability, class, and sexuality. We return to address this important research scarcity in Discussion and Conclusion. 

\section{Methods}

In writing this review article, we have tried to cast as wide a net as possible in seeking out relevant scholarship on the issue of gender equity in physics labs. We have searched through \textit{Physical Review: Physics Education Reearch}, \textit{The Physics Teacher}, the \textit{American Journal of Physics}, the \textit{European Journal of Physics}, \textit{Physics Education}, and the proceedings of the Physics Education Research Conference (PERC), as well as via Google Scholar and with the NCSU library search engine, for combinations of the terms ``gender'' or ``women'' alongside ``physics lab'' or ``physics experiment''.  The NCSU library search engine, QuickSearch, is a customized tool that parses through articles, books, journals, and other resources~\cite{NCSUQuickSearch}. We included publications available up to March, 2023. All publications that were deemed to fit the intent of this review were included, and are cited in the references of this manuscript.

There are six different types of outcome measure used by the studies we review: (1) lab-specific skills and experiences using instruments like ECLASS, (2) indicators of conceptual understanding such as grades and concept inventory scores, (3) academic measures such as passing rates and degree persistence, (4) motivational factors such as self-efficacy and agency, assessed via validated surveys, (5) measures of task division and peer interactions, and (6) qualitative outcomes such as students describing their experiences in lab courses. There is the potential of unexamined gender bias in these measures~\cite{MalespinaGradePenalty}, which have not been evaluated for bias in the same way as some more commonly used instruments~\cite{TraxlerGenderFCI}.

Our intention is to provide an overview of research that has addressed the issue of gender in physics labs up to this point, not to write a meta-analysis. However, in order to provide some structure to this paper, we have tried to draw together some overarching themes from the reviewed scholarship in the Discussion and Conclusion section of this paper. We cannot claim that the references here are a complete representation of research relating to gender in physics labs, but we believe they provide as comprehensive and thorough an overview as possible.

We acknowledge that our positionality played a role in the process of searching out, reading, and analyzing the papers described below~\cite{MilnerPositionality}. Danny is a white man who teaches physics, a discipline that pays him unearned `wages'~\cite{RoedigerWages} because of his identities. His 2021 Ph.D. thesis (with Chandralekha) focused on gender equity in labs, and his involvement in physics education organizations led to conversations with other scholars interested in gender equity in physics labs, many of whose work is included in this review paper. Chandralekha is an Asian woman who has worked with Danny to make physics lab learning environments equitable and inclusive.

We feel that it is important to distinguish between scholarship that is primarily focused with identifying and elucidating gender differences in physics labs on one hand, and evaluation of proposals intended to remedy gender inequities on the other hand. For that reason, the remainder of the paper is divided into three sections. In Section III, we review papers that are primarily concerned with the uncovering of the existence and causes of gender differences in physics labs. In Section IV, we review papers that present and evaluate proposals to reduce the effect of inequities in physics labs. In Section V, we attempt to pull together some overarching themes from the papers and present our vision of avenues for future research.

\section{Manifestations and Causes}

Before reviewing the range of studies of gender in physics labs, it may be useful to look at three `early' studies from the early 2000s. These studies are informative because they illustrate the scope, as well as some of the shortcomings, of research that has thus far been conducted on gender in physics labs.

As part of a Ph.D. dissertation completed in 1997, Kostas analyzed a variety of lab manuals from secondary school physics and other disciplines. Hypothesizing that different types of experimental work might provide different benefits for girls than for boys, Kostas found that the lab manuals in fact provided a balanced combinations of different types of experimental work~\cite{KostasLabManuals}. This null result highlights the challenge of determining causal factors in the physics lab that can be identified as primary causes for gender differences. 
This result also illustrates the importance of starting with a reliable gender framework while designing research projects, rather than relying on assumptions rooted in biological essentialism. As argued in papers referenced later, it is instead likely that gender differences in outcomes arise in a complex way from an entanglement of social, cultural, motivational, and pedagogical factors.

In a 1998 paper, Brown, Slater, and Adams noted that the women (there were 28) in their class of 221 were more likely than the men to use a ``naive method'' when prompted to connect an electric circuit~\cite{BrownBatteriesBulbs}. This finding is valuable, they claim, because ``we are often at a loss to identify specific things we do in our classrooms that place females at a disadvantage''~\cite{BrownBatteriesBulbs}. This research is built on the troublesome assumptions that gender differences are inherent rather than primarily social in nature, and that physics instruction is (and was) equitable for men and women. In a letter to the editor responding to the Brown, Slater, and Adams paper, Wainwright attributed the gender difference to anxiety and self-confidence and shared pedagogical approaches that can help all students learn to build electric circuits effectively~\cite{WainwrightBrownResponse}. The response by Wainwright illustrates how gender-focused studies built on a student deficit framing~\cite{CotnerDeficit,ExarhosDeficitFraming,NairDeficitFraming} like Brown, Slater, and Adams tend to overlook the role of underlying causes while providing little actionable guidance for instruction. This study and the previous should be seen as examples of how \textit{not} to conduct gender-focused scholarship.

A 2001 PERC paper by Potter et al. claimed that the transformation of an introductory physics course to a studio format helped to improve gender parity in final grade distributions~\cite{PotterReformedCourse} compared with a traditional format. However, in a follow-up publication in 2004, the authors expressed continued uncertainty about the mechanisms by which a switch to active learning may have impacted the social component of learning~\cite{McKinnonLabInstructions}. This scholarship anticipated research results that were generally positive, if uneven and hard to qualify, focused on studio-style courses.

\subsection{Surveys}

A number of validated surveys have been developed for use in physics lab courses, including the Physics Measurement Questionnaire (PMQ)~\cite{WolkwynPMQ}, the Concise Data Processing Assessment (CDPA)~\cite{DayCDPA}, the Colorado Learning Attitudes about Science Survey for experimental physics (ECLASS)~\cite{WilcoxECLASS}, the Physics Lab Inventory of Critical thinking (PLIC)~\cite{WalshPLIC}, and the Survey of Physics Reasoning on Uncertainty Concepts in Experiments (SPRUCE)~\cite{VignalSPRUCE}. Data from several of these surveys have been analyzed in order to identify binary gendered trends in student responses. Day et al. found significant gender gaps in performance on the CDPA that persisted from pre-test to post-test~\cite{DayGenderLabs}.

In an analysis of data from the ECLASS, Wilcox and Lewandowski found similar aggregate gender differences that persisted from pre-test to post-test. However, when they disaggreated their student population, the gender differences disappeared for some categories of students (e.g, physics majors in beyond first year lab courses) and persisted for others (e.g., physics non-majors in first year lab courses)~\cite{WilcoxECLASSgender}.

\subsection{Lab Curricula}

The ECLASS and PLIC have both been used to analyze gender differences in labs with different curricula. For example, Doucette et al. reported gender differences in post ECLASS scores for a conceptual inquiry-based introductory physics lab course~\cite{DoucetteCJP}.

In another analysis, Wilcox and Lewandowski found that post-instruction ECLASS scores in skills-based labs eliminated (or even reversed, in specific cases) the pre-instruction `gender gap'~\cite{WilcoxECLASSSkills}. A study by Walsh et al. compared both ECLASS and PLIC results, finding that students of all genders had higher post-instruction scores in skills-based labs than in concepts-focused labs~\cite{WalshSkillsLabs}, controling for pre-instruction scores. A study by Sulaiman et al., focusing on a lab transformation to a skills-based approach, identified some ECLASS items that had different responses from men than for women, suggesting that gender identity can play a role in how students perceive their lab~\cite{SulaimanECLASSgender2}.

\subsection{Integrated Labs}

Introductory physics courses that integrate lecture and lab components together, also known as studio physics, have also been analyzed with an eye on gender. Kohl and Kuo reported that a transition to studio physics narrowed gender differences in course grades, DFW rates and, in particular, scores on a concept inventory of electricity and magnetism concepts~\cite{KohlStudio}. Potter et al. reported eliminating gender differences in grade outcomes after transitioning to a studio-style approach, attributing the evening out of grades to the features of the transformed course~\cite{PotterReformedCourse}. Laws et al. reported that women in a workshop physics class were equally likely as men to choose to major in physics~\cite{LawsWorkshopPhysicsWomensResponses}.

Analysis by Beichner et al. indicated that the SCALE-UP approach was associated with significant reduction in failure rates for integrated introductory physics classes, with the reduction for women outpacing the reduction for men~\cite{BeichnerSCALEUP}. Meanwhile, Brahmia demonstrated that the ISLE approach combined with mathematical reasoning skill-developing activities in a specially-designed course is associated with an increase in the passing rate of a first-year physics course, eliminating earlier gender differences~\cite{BrahmiaImprovingLearning}. 

Focusing on university-level Modeling Instruction, analysis by Traxler and Brewe found equitable outcomes for men and women on an attitudinal survey~\cite{TraxlerModeling}, with no gender-inequitable impacts on long-term measures such as persistence to degree or failure of upper division classes~\cite{RodriguezGenderReformed}. However, there was an increase in the `gender gap' as measured by a concept inventory, which increased between pre and post~\cite{BreweTowardEquity} for those university-level Modeling Instruction classes, with no clear dependence on the instructor's gender~\cite{McPaddenMIGender}.

\subsection{Online Labs}

Prior to the adoption of remote instruction as a result of the COVID-19 pandemic starting in 2020, Kepple analyzed student attitudes and performance in online physics labs using the IOLab device. Kepple reported gains on a concept inventory for both men and women (although the gains for men were greater), and reported that both men and women achieved higher grades in the online lab and also that both men and women would prefer to take an online lab if they had an option to do so~\cite{KeppleDistance}. A study by Rosen and Kelly, focused on socialization, physics epistemology, and help-seeking beliefs, found no gender difference between men and women in the cases of both online and in-person labs~\cite{RosenEpistemology}, also using the IOLab device. Further, they suggest that self-selection into either online or in-person labs may be responsible for reducing `gender gaps' in these factors (implying that perhaps gender gaps may previously have emerged because of students being required to learn in unfavorable contexts).

Based on data collected as a result of the switch to remote instruction as a result of the COVID-19 pandemic, Radulovi{\'c} et al. suggested that online labs may have been somewhat more effective than in-person labs at helping female students learn physics concepts and develop a (self-reported) sense of confidence about conducting experiments~\cite{RadulovicVirtualMotivation}. 
Furthermore, this study showed no difference between male and female students on a separate measure of self-efficacy, which suggests that gender differences in some motivational factors may be partially affected by lab format.

\subsection{High School Labs}

Three studies lend insight into the impact of high school physics labs on gender inequities. Using a national dataset for the USA, Burkam et al. found that hands-on lab activities (which they suggest to be relatively rare) had a positive impact on students' performance, with an especially strong impact for girls~\cite{BurkamHighSchool}. Based on another national dataset for the USA, Hazari et al. reported that, among other things, male high school students were more likely to report that labs addressed beliefs they held about the world~\cite{HazariGenderDiffTPT}. 

Meanwhile, Stoeckel's analysis of one AP physics class identified labs as being especially important for the development of students' sense of confidence, which has been implicated in gender differences~\cite{StoeckelConfidence}. 

\subsection{Upper Division Labs}

In her Ph.D. thesis, Thomas used quantitative and qualitative methods to explore the impact of a novel series of lab experiences that supplement instruction for most courses in the traditional physics major program. This scholarship identified factors in the design of these lab experiences as being essential to the development of students' senses of belonging and identification with physics, as well as their persistence~\cite{ThomasPhD}. Observing gender dynamics in an intermediate astrophysics lab course, Gunter et al. noted that a lack of meaningful discussions about physics concepts, as well as division of labor in which men tended to adopt the hands-on activities, led to gaps in students' understanding. Notably, they also observed that women who were physics majors were viewed as competent and took on leadership roles in groups in which they were paired with men who were not physics majors~\cite{GunterAstroLab}. 

\subsection{Masculinities in Research Labs}

In an ethnographic study of physics research labs in the USA and Japan, Traweek noted the impact of gender and masculinity on the physics research being conducted~\cite{TraweekBeamtimes}. Beyond their focus on scientific exploration, research labs are also a space for education, which inspired three parallel streams of scholarship by Gonsalves, Danielsson, and Pettersson, who used ethnographic approaches to understand the impact of different aspects of masculinity on research labs in the USA, Canada, and Sweden~\cite{GonsalvesMasculinities}. Pettersson described the role of a `boys and their toys' attitude, as well as the significance of large machines and opportunities for students to get their hands dirty, as important factors in a plasma physics lab~\cite{PetterssonMaking,GonsalvesMasculinities}. Gonsalves noted that devices are often constructed with in-built assumptions about the physical bodies (arm length, physical strength, manual dexterity) that will operate them. She also found that women often engage in gender negotiation in physics lab spaces, such as rejecting traditional gender norms or seeking out unique ways to gain recognition for skilled work~\cite{GonsalvesMasculinities}. Danielsson's interviews revealed the significance of tinkering and `female masculinity' in the development of students' identity as a physicist~\cite{DanielssonDoing,DanielssonExploring,GonsalvesMasculinities}. 

\subsection{Motivational Factors}

A wide variety of studies have investigated the impact of physics labs on gender differences in motivational constructs. Kalender et al. explored the role of agency (``the capacity to
guide one’s actions towards achieving a goal'') in labs, finding a lower level of agency for women in a physics majors' lab but not in an engineering majors's lab course~\cite{KalenderAgencyLabs,KalenderAgencyPERC}. Kinnischtzke and Smith noted a tight relationship between agency and self-efficacy in physics labs~\cite{KinnischtzkeAgencySE}. Laws et al. found that women were less confident than men about their lab skills in a studio physics class~\cite{LawsWorkshopPhysicsWomensResponses}. Likewise, Stoeckel's work in a high school classroom illustrated the importance of hands-on lab work for developing students' confidence~\cite{StoeckelConfidence}.

Stump and Holmes found that students, and especially men, often do not recognize `managing the group progress' as part of doing physics, which might be problematic as managerial roles are often adopted by women in mixed-gender groups~\cite{StumpRecog}. Meanwhile, Doucette and Singh described a physics student whose identity as a scientist was boosted by the recognition she received as a member of a physical chemistry research group~\cite{DoucetteReflectionsTwoWomen}. Perceived recognition can be a key factor that impacts students' self-efficacy and physics identity~\cite{LiImpact}.

Gunawan et al. explored the role of creativity in a virtual physics lab for high school students~\cite{GunawanCreativity}. Other investigations into motivational factors in lab, such as Funkhouser's work on identity in a transformed college physics lab~\cite{FunkhouserPhD} and Lassen et al.'s work on project ownership~\cite{LassenOwnership}, have not addressed the gender dimension directly but still shine light indirectly on the role of gender in physics lab learning.

\subsection{Student Interactions}

Social interactions between students can be an important way in which gender differences can influence students' motivational and learning outcomes. Jeon et al. explored the role of student interactions and `inchargeness' in group work, finding that gender equity in group-work could not be simply predicted by these factors~\cite{JeonInchargeness}. Investigations using social network analysis by Sundstrom et al. indicated that student demographics and course structure impacted the interactions students had in their lab courses, and also that in some cases students may have been more likely to interact with same-gender peers~\cite{SundstromNetworkFormation}. In another study focusing on a reformed lab, Sundstrom et al. found that students from all-male groups were more likely to interact with peers from other groups than were students from mixed-gender or all-female groups~\cite{SundstromIntergroupNetworks}. 

In a study of a workshop physics program, Laws et al. reported troubling social dynamics: ``women complained of domineering partners, clashes in temperament, being subjected to ridicule, fears that their partners didn't respect them, and feelings that their partners understood far more than they''~\cite{LassenOwnership}. Similar issues emerged in a study by Doucette and Singh, which documented women describing being ignored, talked over, and not taken seriously in physics labs~\cite{DoucetteGroupwork}.

\subsection{Task Division}

One prominent way in which student interactions have been observed to result in gender differences is through task division. Holmes et al. presented data suggesting that men in mixed gender pairs may be more likely than women to use lab equipment~\cite{HolmesGenderLabs}. Quinn et al. found that gendered task division was particularly pronounced in less-structured inquiry-style labs~\cite{QuinnGenderRoles,QuinnPERC}. Dew et al. found that task division was reduced when it came to equipment handling for remote labs in which students constructed their own apparatus~\cite{DewSoUnfair}. Van Domelen investigated gender differences in perceptions of technology use in a lab setting~\cite{VanDomelenGenderEffects}, finding subtle differences between men and women. Akubo et al. presented two case studies in which positioning/power dynamics influenced the ways that students self-assigned different aspects of their lab-work~\cite{AkuboRoleNegotiation}.

Doucette et al. used ethnographic observations to propose that in mixed gender groups, women may be pushed toward secretarial and managerial ('Hermione') work while men are inclined toward tinkering and wasting their time on cellphones and other distractions not related to the physics lab~\cite{DoucetteHermione}. Based upon individual interviews and ethnographic classroom observations of mixed-gender and same-gender groups from the beginning of the semester when the groups first formed, they proposed a model for how the gendered roles solidified in many mixed-gender groups compared to the same-gender groups~\cite{DoucetteHermione}. In this model based upon typical gender dynamics, a student's initiative—their willingness to do work—in the lab is plotted horizontally, while the vertical axis shows a student's gender~\cite{DoucetteHermione}. The model shows that typically when a high-initiative woman begins to work with a low-initiative man she adopts a Hermione role and he becomes a person who is often distracted from the lab work unless she insists that he does certain tasks. On the other hand, when a woman with lower initiative begins to work with a high-initiative man: he tends increasingly to take over the experiment, adopting the Tinkerer role, and she tends more toward the Secretary role. The type of gender dynamics that drives this task division is observed throughout the lab period, but is especially pronounced during the first hour that a pair of students is beginning to work together. Their observations suggest that unlike in mixed-gender groups, the `phase separation' into different roles generally does not seem to occur in same-gender groups. In fact, in their observations, the general contrast between the mixed-gender and same-gender groups in this regard was striking~\cite{DoucetteHermione}.

Studies of women who take on a leadership role in collaborative lab-work have shown that this role can have some positive impact on their identity development~\cite{StumpManager,Doucette4WomenEJP}. For example, in one study, women who were group leaders described that they felt valued as leaders and liked the lab much more than the physics lectures since their team would not successfully complete the lab work without their leadership role~\cite{Doucette4WomenEJP}. However, it is worrisome that many of them conveyed low overall physics self-efficacy and the belief that their male peers were better at tinkering in the lab or physics in general, which may at least partly be due to stereotypes about who can excel in physics. For example, one female lab group leader whose male peer routinely dominated tinkering with the equipment stated, ``He's very good at equipment, so even if he doesn't necessarily read the lab, he's just one of those people that has very good problem-solving skills when there's hands-on things''~\cite{Doucette4WomenEJP}. Critically, as shown by Holmes et al., the emergence of gender-based inequities such as task division cannot be attributed to gender differences in student preferences about collaborative work~\cite{HolmesStudentPreference}.

\section{Proposals to reduce inequities}

Whether we consider the role of gender-based task division, the impact of masculinities in physics labs, or the reasons that students with different gender identities achieve different outcomes in labs, research studies presented here suggest that a central underlying cause of gender-based inequity is the masculine culture of physics. Prior scholarship has proposed or evaluated the success of different approaches to reducing gender-based inequities in physics labs. A wide variety of approaches has been suggested~\cite{BlueResourceLetter}, including the use of course-based undergraduate research experiences~\cite{WerthCURE}, explicitly discussing equity issues~\cite{LockDiscussing}, and social-belonging interventions~\cite{BinningChanging}. In this section, we report only on scholarship focused specifically on studies that have examined strategies for reducing inequities in physics labs.

\subsection{Reducing Isolation}

In a physics lab, women may become isolated in two different ways: they may be excluded if students are allowed to choose their own groups~\cite{DoucetteArduinoLessons,LiImpact}, or they may find that their group-mates do not value their expertise or ignore their contributions~\cite{DoucetteGroupwork}. Past scholarship (e.g., Gillibrand's study of single sex classes in mixed schools~\cite{GillibrandSingleSexClasses}) suggests that one solution to reducing isolation is by ensuring that groups of 3 or more students are formed in such a way as to include either 0 or 2 or more women; in other words, avoiding forming groups with exactly 1 woman. In this physics context, this suggestion appears to originate from Heller et al.~\cite{HellerGroupwork}, and was subsequently incorporated in other curricular projects (e.g., SCALE-UP~\cite{BeichnerSCALEUP}) and scholarship (e.g., McCullough's column in The Physics Teacher~\cite{McCulloughGenderClassroom}).

However, some students may find a `no isolated woman' group formation strategy to be disconcerting. Instructors may instead choose to focus on ways that they can structure and support group-work in ways that helps students develop equitable and fair collaboration practices, such as via group-work contracts~\cite{BrannenGroupworkContract}, interventions~\cite{LewisIntervention}, and making group-work an assessed course outcome~\cite{AAPTlab}. As an example of how the gender composition of groups may not play a significant role for well-structured work, Callan et al. reported that in their studio-style physics classes there were no differences in students' grades, conceptual understanding, or experimental beliefs when comparing mixed-gender groups with single-gender groups~\cite{CallanGroups}.

\subsection{Assigned Group Roles}

If students are inclined to divide up group-work tasks, one strategy that might improve equity is to explicitly assign, and then rotate, group roles. For example, Heller and Heller devised a scheme by which students would rotate through the roles of manager, recorder, skeptic, and summarizer in cooperative problem-solving and lab-work~\cite{HellerCooperative}. Students may be asked to reflect on their role participation, in order to show students that group-work roles are important~(e.g., \cite{DoucetteGroupwork}).

While roles may offer a way for students to participate in all aspects of the lab-work, Holmes et al. have reported that students prefer to share, rather than split up, their lab-work~\cite{HolmesStudentPreference}. Furthermore, sharing rather than splitting up group work can serve to boost the self efficacy of women in mixed gender groups~\cite{DoucetteGoodLabPartnerPERC,DoucetteShareIt}.

\subsection{Attending to the Intersubjective Space}

In a case study of a particular physics department, Johnson described efforts by faculty to provide structural support for the development of students' physics identity. These efforts included teaching students how to collaborate in their lab-work, using evidence-informed instruction, and communicating a growth mindset to the students~\cite{JohnsonIntersectionalFramework}.

Indeed, instructors can have a powerful impact on the experiences that serve to shape their students' physics identities. A strong body of research (e.g.,~\cite{MarshmanSE,KalenderIdentity,HazariIdentity}) has drawn connections between the masculine culture of physics, students' motivational characteristics such as physics identity and self-efficacy, and gender-based inequities in physics outcomes. A key aspect of this connection is the intersubjective space; that is, the space of relationships between people that includes concerns (e.g., `I wonder what he thinks about my physics ability') that can strongly impact students' beliefs about themselves and their abilities. Thus, it has been suggested that instructors might be able to indirectly impact the motivational characteristics of students by addressing the intersubjective space. An intersubjective space that is more focused on collaboration, caring, and hard work can, as Johnson notes, help to lift stereotype threat that women often face in physics settings~\cite{JohnsonIntersectionalFramework}.

For example, Doucette and Singh described how two women who struggled in physics and physical chemistry labs were able to persist and succeed by seeking out recognition and mentorship from beyond their peer groups~\cite{DoucetteReflectionsTwoWomen}. Gosling described the positive impact of providing choice to high school students for a 5-week lab project~\cite{GoslingLabChoice}. The intersubjective space can also be addressed through ecological interventions, such as the ecological belonging intervention pioneered by Binning et al.~\cite{BinningChanging}.

\subsection{Preparing Student Instructors}

Particularly in large research-focused universities, physics labs are often taught by graduate or undergraduate students who may have little pedagogical training. As early as 1940, Sister Ambrosia advocated for an approach to lab instruction that would improve learning outcomes for women in mixed gender physics labs by giving the students opportunities to serve as ``partner-instructors'' to classmates who are about to start an experiment that they had previously completed, themselves~\cite{AmbrosiaTeachingPhysicsToWomen}. Turpen et al. explored how undergraduate learning assistants diagnose and respond to inequitable teamwork in a project-based engineering course~\cite{TurpenGroupInequity}. Several efforts to improve and transform the preparation of graduate student lab instructors have included teaching practices focused on responding to inequitable student groupwork~\cite{MunozTAs,WanTAs,DoucettePDforLabTAs}.

\subsection{Does Studio Physics Reduce Gender Inequities?}

While evidence-based active learning instruction generally produces better academic outcomes on average~\cite{FreemanActiveLearning}, the impact of active learning instruction on gender differences is less clear. Some studies suggest that active learning can reduce or eliminate `gender gaps' in grades and motivational outcomes (e.g.,~\cite{EspinosaGenderGap,LorenzoCrouchMazur,GoodGenderProblemSolving}), while other studies suggest that active learning instruction may preserve or even expand `gender gaps'~\cite{KarimEBAEgender,MariesStereotypeThreat,MariesAgreeing}.

As a form of highly engaging active learning, studio physics might be expected to provide some answers to the question of whether active learning instruction can reliably reduce gender differences. For example, an expanded introductory physics course closed `gender gaps' in final grades~\cite{EtkinaLessonsLearned}, and a transformed studio physics course saw the `gender gap' in final grades disappear~\cite{PotterReformedCourse}. On the other hand, an investigation of a studio physics course by Laws et al. reported that some women found the course unpleasant and also noted that while there were no gender differences in grades, women were less confident about their laboratory skills than men~\cite{LawsWorkshopPhysicsWomensResponses}. The other studies described in Section IIC similarly shed light on this question. 

Likely, as these studies suggest, what matters for gender equity in active learning and physics labs is not the curriculum by itself, but the structure of student interactions, the nature of grading and feedback, the ways in which instructors interact and model behavior for students, and many other factors. Thus, integrated or studio physics may be more or less equitable, depending on details of the course design and implementation.

\section{Discussion and Conclusions}

\subsection{Research Results}

A core research result that emerges across the scholarship we reviewed is the salience and persistence of gender-based inequities in educational physics labs. The different ways that students practice gender have a dramatic and important impact on their perceptions of experimental science, on skill-based and conceptual outcomes, on students' experiences in lab courses, on students' views of themselves as physicists, and on the work they perform in labs. The central outcomes parallel results from the rich tradition of research in lecture-based physics courses~\cite{TraxlerEnriching,BlueResourceLetter}, which has likewise identified gender differences in outcomes, experiences, and attitudes. Many different aspects of the physics lab, as an educational setting, serve to subvert the goal of equity of parity and deprive students (especially women and gender minorities) of opportunities to learn to their full capacity and to come to see themselves as capable of doing experimental physics work.

Several other key ideas emerge from a consideration of the scholarship described in this paper. First, while different lab curricula may have an impact on gender differences in lab-related skills, conceptual understanding, grades, and motivational factors, the success of lab curriculum transformations appears to depend heavily on the details: the structure of student work and interactions, grading policies and expectations, the role of lab instructors, and so on. This suggests that efforts to transform lab instruction to improve equity will hinge on reconciling the insight and institutional knowledge of those who have intimate knowledge of how a particular lab works (e.g., lab coordinators) with the perspective and experience of experts on gender-based effects in educational spaces (e.g., researchers). Instead of searching for a panacea, we should focus on deliberate, sustained, evidence-based improvements to our physics labs in order to improve gender equity.

Second, there are subtle yet important factors associated with different lab environments (high school, upper division, online) and curricula (skills-based, conceptual inquiry, studio physics) that current scholarship suggests may be important. Findings that are relevant at the introductory college level may not translate to high school; for example, a grouping strategy that seems to improve gender-based equity in an introductory college physics lab context may not have the same effect in a high school classroom.

Third, student interactions such as task division appear to play an important role in how gender impacts students' experiences in, and beliefs about, physics labs. Lab instructors should (be taught to) look for inequitable student interactions, and should (be taught to) respond effectively to these interactions.

Fourth, physics lab spaces, and especially research spaces, are often masculinized, which can pose challenges especially for people whose gender identities do not align with masculinity. Particularly in research labs that train physics majors and graduate students, physicists should consider how the work required of students and the climate of the lab may be gendered in physical, ideational, or intersubjective ways.

Fifth, while some strategies to reduce gender inequities in labs have been advanced and investigated, there is a substantial need for more creativity, insight, and scholarship applied to addressing the problem of gender inequity in physics lab spaces.

\subsection{Limitations and Future Directions}

One limitation to this overview stems from the lack of statistical rigor available in the referenced studies. Few of the studies provide a quantitative analysis that includes effect sizes, for example. Further, when effect sizes are either included or can be determined, they often disagree: Cohen's d for women vs. men in a first-year lab is -0.4 in one study~\cite{WilcoxECLASSgender} and -0.1 in another~\cite{DoucetteCJP}. As yet, there has not be satisfactory quantitative replication of broad gendered patterns across institutions and contexts.

It is important to note not only what scholarship has been conducted, but also what dimensions of physics lab instruction have not been put under the scrutiny of a gender equity analysis. As with much of the field of physics education research~\cite{KanimDemographics}, introductory college physics labs at large research-focused universities have been studied more extensively than physics labs at high schools and two-year colleges, labs at minority-serving institutions, `beyond the first year' and graduate level physics labs, or physics lab courses outside the USA. Similarly, much focus has been applied to gender differences, but there has been very little scholarship about the effectiveness of different strategies for improving gender equity in physics labs.

We found no studies that specifically reported on the physics lab experiences of students with non-binary gender identities. Most of the studies come from universities in the USA, and large research universities are strongly over-represented. Thus, we urge the reader to be cautious to avoid over-generalizing the results of any one study, or even our holistic analysis, beyond the study's original context.

In some cases, the focus on gender differences in the scholarship described previously has had the effect of collapsing the complex, multi-faceted issue of gender inequity into a simple binary comparison. There is value in comparing the outcomes of men and women in lab settings, for example comparing sense of belonging for men and women, especially when studies can be conducted with large populations that allow for the use of model-based analysis (such as structural equation modeling or network analysis). However, it is important that such studies avoid implicit assumptions in the framing of research questions (e.g., student deficit framing~\cite{CotnerDeficit}) and in drawing conclusions based on comparative results.

However, it is essential that gender-focused research also addresses ways in which gender identity transcends fixed, binary categorization. There is a need for more scholarship that investigates the experiences and outcomes for students with a more sophisticated and complex gender framing. Future scholarship will need to be thoughtful and imaginative about employing frameworks such as masculinity that provide a foundation for researchers to analyze deeply the causes and mechanisms that underlie their observations.

While there has been some scholarship focused on gender in physics labs, very little focus has been directed toward other aspects of identity. More than anything else, we recommend that future scholarship on equity in physics labs adopts intersectional analyses that also account for race, ethnicity, disability status, class, sexuality, and other dimensions of identity. A report written by Dounas-Frazer et al. for the AAPT Committee on Laboratories that calls for the investment of time, energy, and resources to increase accessibility for physics labs~\cite{DounasFrazerAccessibleLabs} is an important first step in this direction. Likewise, the intersectional (gender and race) analysis by Akubo at al. of the negotiation of group roles for students in small lab groups~\cite{AkuboRoleNegotiation} is an important benchmark for future intersectional work.

We encourage researchers to pursue scholarship that investigates students' experiences and outcomes in high school physics labs, in two year college physics labs, in intermediate and advanced undergraduate lab courses, and in physics lab courses both in and outside of the USA at different types of academic institutions. We also encourage researchers to conduct research that focuses on the experiences of students who navigate gender in non-traditional ways, as well as research that investigates the impact of other axes of identity, as well as intersectionality of identities, on students' experiences and outcomes in physics labs.

\bibliography{the}

\begin{thebibliography}{120}%
\makeatletter
\providecommand \@ifxundefined [1]{%
 \@ifx{#1\undefined}
}%
\providecommand \@ifnum [1]{%
 \ifnum #1\expandafter \@firstoftwo
 \else \expandafter \@secondoftwo
 \fi
}%
\providecommand \@ifx [1]{%
 \ifx #1\expandafter \@firstoftwo
 \else \expandafter \@secondoftwo
 \fi
}%
\providecommand \natexlab [1]{#1}%
\providecommand \enquote  [1]{``#1''}%
\providecommand \bibnamefont  [1]{#1}%
\providecommand \bibfnamefont [1]{#1}%
\providecommand \citenamefont [1]{#1}%
\providecommand \href@noop [0]{\@secondoftwo}%
\providecommand \href [0]{\begingroup \@sanitize@url \@href}%
\providecommand \@href[1]{\@@startlink{#1}\@@href}%
\providecommand \@@href[1]{\endgroup#1\@@endlink}%
\providecommand \@sanitize@url [0]{\catcode `\\12\catcode `\$12\catcode
  `\&12\catcode `\#12\catcode `\^12\catcode `\_12\catcode `\%12\relax}%
\providecommand \@@startlink[1]{}%
\providecommand \@@endlink[0]{}%
\providecommand \url  [0]{\begingroup\@sanitize@url \@url }%
\providecommand \@url [1]{\endgroup\@href {#1}{\urlprefix }}%
\providecommand \urlprefix  [0]{URL }%
\providecommand \Eprint [0]{\href }%
\providecommand \doibase [0]{https://doi.org/}%
\providecommand \selectlanguage [0]{\@gobble}%
\providecommand \bibinfo  [0]{\@secondoftwo}%
\providecommand \bibfield  [0]{\@secondoftwo}%
\providecommand \translation [1]{[#1]}%
\providecommand \BibitemOpen [0]{}%
\providecommand \bibitemStop [0]{}%
\providecommand \bibitemNoStop [0]{.\EOS\space}%
\providecommand \EOS [0]{\spacefactor3000\relax}%
\providecommand \BibitemShut  [1]{\csname bibitem#1\endcsname}%
\let\auto@bib@innerbib\@empty
\bibitem [{\citenamefont {DeBoer}(1991)}]{DeboerHistory}%
  \BibitemOpen
  \bibfield  {author} {\bibinfo {author} {\bibfnamefont {G.}~\bibnamefont
  {DeBoer}},\ }\href@noop {} {\emph {\bibinfo {title} {A History of Ideas in
  Science Education}}}\ (\bibinfo  {publisher} {Teachers College Press},\
  \bibinfo {year} {1991})\BibitemShut {NoStop}%
\bibitem [{\citenamefont {Cwik}\ and\ \citenamefont
  {Singh}(2023)}]{CwikFramework}%
  \BibitemOpen
  \bibfield  {author} {\bibinfo {author} {\bibfnamefont {S.}~\bibnamefont
  {Cwik}}\ and\ \bibinfo {author} {\bibfnamefont {C.}~\bibnamefont {Singh}},\
  }\bibfield  {title} {\bibinfo {title} {Framework for and review of research
  on assessing and improving equity and inclusion in undergraduate physics
  learning environments},\ }in\ \href
  {https://doi.org/https://doi.org/10.1063/9780735425514_002} {\emph {\bibinfo
  {booktitle} {International Handbook of Physics Education Research: Special
  Topics}}},\ \bibinfo {editor} {edited by\ \bibinfo {editor} {\bibfnamefont
  {M.~F.}\ \bibnamefont {Taşar}}\ and\ \bibinfo {editor} {\bibfnamefont
  {P.~R.~L.}\ \bibnamefont {Heron}}}\ (\bibinfo  {publisher} {AIP Publishing,
  Melville, New York},\ \bibinfo {year} {2023})\ pp.\ \bibinfo {pages}
  {2--1---2--26}\BibitemShut {NoStop}%
\bibitem [{\citenamefont {Rodriguez}\ \emph {et~al.}(2012)\citenamefont
  {Rodriguez}, \citenamefont {Brewe}, \citenamefont {Sawtelle},\ and\
  \citenamefont {Kramer}}]{RodriguezEquityModels}%
  \BibitemOpen
  \bibfield  {author} {\bibinfo {author} {\bibfnamefont {I.}~\bibnamefont
  {Rodriguez}}, \bibinfo {author} {\bibfnamefont {E.}~\bibnamefont {Brewe}},
  \bibinfo {author} {\bibfnamefont {V.}~\bibnamefont {Sawtelle}},\ and\
  \bibinfo {author} {\bibfnamefont {L.~H.}\ \bibnamefont {Kramer}},\ }\bibfield
   {title} {\bibinfo {title} {Impact of equity models and statistical measures
  on interpretations of educational reform},\ }\href
  {https://doi.org/10.1103/PhysRevSTPER.8.020103} {\bibfield  {journal}
  {\bibinfo  {journal} {Phys. Rev. ST Phys. Educ. Res.}\ }\textbf {\bibinfo
  {volume} {8}},\ \bibinfo {pages} {020103} (\bibinfo {year}
  {2012})}\BibitemShut {NoStop}%
\bibitem [{\citenamefont {Cochran}(2018)}]{CochranKitchen}%
  \BibitemOpen
  \bibfield  {author} {\bibinfo {author} {\bibfnamefont {G.}~\bibnamefont
  {Cochran}},\ }\bibfield  {title} {\bibinfo {title} {Guest post: The problem
  with diversity, inclusion, and equity},\ }\href@noop {} {\bibfield  {journal}
  {\bibinfo  {journal} {The Scholarly Kitchen}\ }\textbf {\bibinfo {volume}
  {28}} (\bibinfo {year} {2018})}\BibitemShut {NoStop}%
\bibitem [{\citenamefont {Guti{\'e}rrez}(2008)}]{GutierrezGapGazing}%
  \BibitemOpen
  \bibfield  {author} {\bibinfo {author} {\bibfnamefont {R.}~\bibnamefont
  {Guti{\'e}rrez}},\ }\bibfield  {title} {\bibinfo {title} {Research
  commentary: A gap-gazing fetish in mathematics education? problematizing
  research on the achievement gap},\ }\href@noop {} {\bibfield  {journal}
  {\bibinfo  {journal} {Journal for research in mathematics education}\
  }\textbf {\bibinfo {volume} {39}},\ \bibinfo {pages} {357} (\bibinfo {year}
  {2008})}\BibitemShut {NoStop}%
\bibitem [{\citenamefont {Holmes}\ and\ \citenamefont
  {Lewandowski}(2020)}]{HolmesLandscape}%
  \BibitemOpen
  \bibfield  {author} {\bibinfo {author} {\bibfnamefont {N.~G.}\ \bibnamefont
  {Holmes}}\ and\ \bibinfo {author} {\bibfnamefont {H.~J.}\ \bibnamefont
  {Lewandowski}},\ }\bibfield  {title} {\bibinfo {title} {Investigating the
  landscape of physics laboratory instruction across {North America}},\ }\href
  {https://doi.org/10.1103/PhysRevPhysEducRes.16.020162} {\bibfield  {journal}
  {\bibinfo  {journal} {Phys. Rev. Phys. Educ. Res.}\ }\textbf {\bibinfo
  {volume} {16}},\ \bibinfo {pages} {020162} (\bibinfo {year}
  {2020})}\BibitemShut {NoStop}%
\bibitem [{\citenamefont {Steele}(2011)}]{SteeleWhistling}%
  \BibitemOpen
  \bibfield  {author} {\bibinfo {author} {\bibfnamefont {C.~M.}\ \bibnamefont
  {Steele}},\ }\href@noop {} {\emph {\bibinfo {title} {Whistling Vivaldi: How
  stereotypes affect us and what we can do}}}\ (\bibinfo  {publisher} {WW
  Norton \& Company},\ \bibinfo {year} {2011})\BibitemShut {NoStop}%
\bibitem [{\citenamefont {Maries}\ \emph {et~al.}(2018)\citenamefont {Maries},
  \citenamefont {Karim},\ and\ \citenamefont {Singh}}]{MariesAgreeing}%
  \BibitemOpen
  \bibfield  {author} {\bibinfo {author} {\bibfnamefont {A.}~\bibnamefont
  {Maries}}, \bibinfo {author} {\bibfnamefont {N.~I.}\ \bibnamefont {Karim}},\
  and\ \bibinfo {author} {\bibfnamefont {C.}~\bibnamefont {Singh}},\ }\bibfield
   {title} {\bibinfo {title} {Is agreeing with a gender stereotype correlated
  with the performance of female students in introductory physics?},\
  }\href@noop {} {\bibfield  {journal} {\bibinfo  {journal} {Physical Review
  Physics Education Research}\ }\textbf {\bibinfo {volume} {14}},\ \bibinfo
  {pages} {020119} (\bibinfo {year} {2018})}\BibitemShut {NoStop}%
\bibitem [{\citenamefont {Leslie}\ \emph {et~al.}(2015)\citenamefont {Leslie},
  \citenamefont {Cimpian}, \citenamefont {Meyer},\ and\ \citenamefont
  {Freeland}}]{LeslieBrilliance}%
  \BibitemOpen
  \bibfield  {author} {\bibinfo {author} {\bibfnamefont {S.-J.}\ \bibnamefont
  {Leslie}}, \bibinfo {author} {\bibfnamefont {A.}~\bibnamefont {Cimpian}},
  \bibinfo {author} {\bibfnamefont {M.}~\bibnamefont {Meyer}},\ and\ \bibinfo
  {author} {\bibfnamefont {E.}~\bibnamefont {Freeland}},\ }\bibfield  {title}
  {\bibinfo {title} {Expectations of brilliance underlie gender distributions
  across academic disciplines},\ }\href@noop {} {\bibfield  {journal} {\bibinfo
   {journal} {Science}\ }\textbf {\bibinfo {volume} {347}},\ \bibinfo {pages}
  {262} (\bibinfo {year} {2015})}\BibitemShut {NoStop}%
\bibitem [{\citenamefont {Weitekamp}(2017)}]{WeitekampBigBangTheory}%
  \BibitemOpen
  \bibfield  {author} {\bibinfo {author} {\bibfnamefont {M.~A.}\ \bibnamefont
  {Weitekamp}},\ }\bibfield  {title} {\bibinfo {title} {The image of scientists
  in {The Big Bang Theory}},\ }\href@noop {} {\bibfield  {journal} {\bibinfo
  {journal} {Physics Today}\ ,\ \bibinfo {pages} {40}} (\bibinfo {year}
  {2017})}\BibitemShut {NoStop}%
\bibitem [{\citenamefont {Wertheim}(1997)}]{WertheimPythagoras}%
  \BibitemOpen
  \bibfield  {author} {\bibinfo {author} {\bibfnamefont {M.}~\bibnamefont
  {Wertheim}},\ }\href@noop {} {\emph {\bibinfo {title} {Pythagoras' Trousers:
  God, Physics, and the Gender Wars}}}\ (\bibinfo  {publisher} {WW Norton \&
  Company},\ \bibinfo {year} {1997})\BibitemShut {NoStop}%
\bibitem [{\citenamefont {Hasse}(2002)}]{HasseGender}%
  \BibitemOpen
  \bibfield  {author} {\bibinfo {author} {\bibfnamefont {C.}~\bibnamefont
  {Hasse}},\ }\bibfield  {title} {\bibinfo {title} {Gender diversity in play
  with physics: The problem of premises for participation in activities},\
  }\href@noop {} {\bibfield  {journal} {\bibinfo  {journal} {Mind, Culture, and
  Activity}\ }\textbf {\bibinfo {volume} {9}},\ \bibinfo {pages} {250}
  (\bibinfo {year} {2002})}\BibitemShut {NoStop}%
\bibitem [{\citenamefont {Gonsalves}\ \emph {et~al.}(2016)\citenamefont
  {Gonsalves}, \citenamefont {Danielsson},\ and\ \citenamefont
  {Pettersson}}]{GonsalvesMasculinities}%
  \BibitemOpen
  \bibfield  {author} {\bibinfo {author} {\bibfnamefont {A.~J.}\ \bibnamefont
  {Gonsalves}}, \bibinfo {author} {\bibfnamefont {A.}~\bibnamefont
  {Danielsson}},\ and\ \bibinfo {author} {\bibfnamefont {H.}~\bibnamefont
  {Pettersson}},\ }\bibfield  {title} {\bibinfo {title} {Masculinities and
  experimental practices in physics: The view from three case studies},\
  }\href@noop {} {\bibfield  {journal} {\bibinfo  {journal} {Phys. Rev. Phys.
  Educ. Res.}\ }\textbf {\bibinfo {volume} {12}},\ \bibinfo {pages} {020120}
  (\bibinfo {year} {2016})}\BibitemShut {NoStop}%
\bibitem [{\citenamefont {Traxler}\ \emph {et~al.}(2016)\citenamefont
  {Traxler}, \citenamefont {Cid}, \citenamefont {Blue},\ and\ \citenamefont
  {Barthelemy}}]{TraxlerEnriching}%
  \BibitemOpen
  \bibfield  {author} {\bibinfo {author} {\bibfnamefont {A.~L.}\ \bibnamefont
  {Traxler}}, \bibinfo {author} {\bibfnamefont {X.~C.}\ \bibnamefont {Cid}},
  \bibinfo {author} {\bibfnamefont {J.}~\bibnamefont {Blue}},\ and\ \bibinfo
  {author} {\bibfnamefont {R.}~\bibnamefont {Barthelemy}},\ }\bibfield  {title}
  {\bibinfo {title} {Enriching gender in physics education research: A binary
  past and a complex future},\ }\href
  {https://doi.org/10.1103/PhysRevPhysEducRes.12.020114} {\bibfield  {journal}
  {\bibinfo  {journal} {Phys. Rev. Phys. Educ. Res.}\ }\textbf {\bibinfo
  {volume} {12}},\ \bibinfo {pages} {020114} (\bibinfo {year}
  {2016})}\BibitemShut {NoStop}%
\bibitem [{\citenamefont {Danielsson}\ and\ \citenamefont
  {Linder}(2009)}]{DanielssonLearning}%
  \BibitemOpen
  \bibfield  {author} {\bibinfo {author} {\bibfnamefont {A.~T.}\ \bibnamefont
  {Danielsson}}\ and\ \bibinfo {author} {\bibfnamefont {C.}~\bibnamefont
  {Linder}},\ }\bibfield  {title} {\bibinfo {title} {Learning in physics by
  doing laboratory work: Towards a new conceptual framework},\ }\href@noop {}
  {\bibfield  {journal} {\bibinfo  {journal} {Gender and Education}\ }\textbf
  {\bibinfo {volume} {21}},\ \bibinfo {pages} {129} (\bibinfo {year}
  {2009})}\BibitemShut {NoStop}%
\bibitem [{\citenamefont {West}\ and\ \citenamefont
  {Zimmerman}(1987)}]{WestDoingGender}%
  \BibitemOpen
  \bibfield  {author} {\bibinfo {author} {\bibfnamefont {C.}~\bibnamefont
  {West}}\ and\ \bibinfo {author} {\bibfnamefont {D.~H.}\ \bibnamefont
  {Zimmerman}},\ }\bibfield  {title} {\bibinfo {title} {Doing gender},\
  }\href@noop {} {\bibfield  {journal} {\bibinfo  {journal} {Gend. Soc.}\
  }\textbf {\bibinfo {volume} {1}},\ \bibinfo {pages} {125} (\bibinfo {year}
  {1987})}\BibitemShut {NoStop}%
\bibitem [{\citenamefont {Connell}(2005)}]{ConnellMasculinities}%
  \BibitemOpen
  \bibfield  {author} {\bibinfo {author} {\bibfnamefont {R.~W.}\ \bibnamefont
  {Connell}},\ }\href@noop {} {\emph {\bibinfo {title} {Masculinities}}},\
  \bibinfo {edition} {2nd}\ ed.\ (\bibinfo  {publisher} {University of
  California Press},\ \bibinfo {year} {2005})\BibitemShut {NoStop}%
\bibitem [{\citenamefont {Hall}\ and\ \citenamefont
  {Sandler}(1982)}]{HallChillyClimate}%
  \BibitemOpen
  \bibfield  {author} {\bibinfo {author} {\bibfnamefont {R.~M.}\ \bibnamefont
  {Hall}}\ and\ \bibinfo {author} {\bibfnamefont {B.~R.}\ \bibnamefont
  {Sandler}},\ }\href@noop {} {\emph {\bibinfo {title} {The Classroom Climate:
  A Chilly One for Women?}}},\ \bibinfo {type} {Tech. Rep.}\ (\bibinfo
  {institution} {Association of American Colleges},\ \bibinfo {year}
  {1982})\BibitemShut {NoStop}%
\bibitem [{\citenamefont {Seymour}\ and\ \citenamefont
  {Hewitt}(1997)}]{SeymourTalking}%
  \BibitemOpen
  \bibfield  {author} {\bibinfo {author} {\bibfnamefont {E.}~\bibnamefont
  {Seymour}}\ and\ \bibinfo {author} {\bibfnamefont {N.~M.}\ \bibnamefont
  {Hewitt}},\ }\href@noop {} {\emph {\bibinfo {title} {Talking about
  Leaving}}}\ (\bibinfo  {publisher} {Westview Press, Boulder, CO},\ \bibinfo
  {year} {1997})\BibitemShut {NoStop}%
\bibitem [{\citenamefont {Kanim}\ and\ \citenamefont
  {Cid}(2020)}]{KanimDemographics}%
  \BibitemOpen
  \bibfield  {author} {\bibinfo {author} {\bibfnamefont {S.}~\bibnamefont
  {Kanim}}\ and\ \bibinfo {author} {\bibfnamefont {X.~C.}\ \bibnamefont
  {Cid}},\ }\bibfield  {title} {\bibinfo {title} {Demographics of physics
  education research},\ }\href@noop {} {\bibfield  {journal} {\bibinfo
  {journal} {Physical Review Physics Education Research}\ }\textbf {\bibinfo
  {volume} {16}},\ \bibinfo {pages} {020106} (\bibinfo {year}
  {2020})}\BibitemShut {NoStop}%
\bibitem [{\citenamefont {Beswick}\ \emph {et~al.}()\citenamefont {Beswick},
  \citenamefont {Boyer},\ and\ \citenamefont {Lown}}]{NCSUQuickSearch}%
  \BibitemOpen
  \bibfield  {author} {\bibinfo {author} {\bibfnamefont {K.}~\bibnamefont
  {Beswick}}, \bibinfo {author} {\bibfnamefont {J.}~\bibnamefont {Boyer}},\
  and\ \bibinfo {author} {\bibfnamefont {C.}~\bibnamefont {Lown}},\ }\href@noop
  {} {\bibinfo {title} {Quicksearch}},\ \bibinfo {note}
  {https://www.lib.ncsu.edu/projects/quicksearch}\BibitemShut {NoStop}%
\bibitem [{\citenamefont {Malespina}\ and\ \citenamefont
  {Singh}(2022)}]{MalespinaGradePenalty}%
  \BibitemOpen
  \bibfield  {author} {\bibinfo {author} {\bibfnamefont {A.}~\bibnamefont
  {Malespina}}\ and\ \bibinfo {author} {\bibfnamefont {C.}~\bibnamefont
  {Singh}},\ }\bibfield  {title} {\bibinfo {title} {Impact of grade penalty in
  first-year foundational science courses on female engineering majors},\
  }\href@noop {} {\bibfield  {journal} {\bibinfo  {journal} {International
  Journal of Engineering Education}\ }\textbf {\bibinfo {volume} {38}},\
  \bibinfo {pages} {1021} (\bibinfo {year} {2022})}\BibitemShut {NoStop}%
\bibitem [{\citenamefont {Traxler}\ \emph {et~al.}(2018)\citenamefont
  {Traxler}, \citenamefont {Henderson}, \citenamefont {Stewart}, \citenamefont
  {Stewart}, \citenamefont {Papak},\ and\ \citenamefont
  {Lindell}}]{TraxlerGenderFCI}%
  \BibitemOpen
  \bibfield  {author} {\bibinfo {author} {\bibfnamefont {A.}~\bibnamefont
  {Traxler}}, \bibinfo {author} {\bibfnamefont {R.}~\bibnamefont {Henderson}},
  \bibinfo {author} {\bibfnamefont {J.}~\bibnamefont {Stewart}}, \bibinfo
  {author} {\bibfnamefont {G.}~\bibnamefont {Stewart}}, \bibinfo {author}
  {\bibfnamefont {A.}~\bibnamefont {Papak}},\ and\ \bibinfo {author}
  {\bibfnamefont {R.}~\bibnamefont {Lindell}},\ }\bibfield  {title} {\bibinfo
  {title} {Gender fairness within the force concept inventory},\ }\href
  {https://doi.org/10.1103/PhysRevPhysEducRes.14.010103} {\bibfield  {journal}
  {\bibinfo  {journal} {Phys. Rev. Phys. Educ. Res.}\ }\textbf {\bibinfo
  {volume} {14}},\ \bibinfo {pages} {010103} (\bibinfo {year}
  {2018})}\BibitemShut {NoStop}%
\bibitem [{\citenamefont {Milner~IV}(2007)}]{MilnerPositionality}%
  \BibitemOpen
  \bibfield  {author} {\bibinfo {author} {\bibfnamefont {H.~R.}\ \bibnamefont
  {Milner~IV}},\ }\bibfield  {title} {\bibinfo {title} {Race, culture, and
  researcher positionality: Working through dangers seen, unseen, and
  unforeseen},\ }\href@noop {} {\bibfield  {journal} {\bibinfo  {journal}
  {Educational Researcher}\ }\textbf {\bibinfo {volume} {36}},\ \bibinfo
  {pages} {388} (\bibinfo {year} {2007})}\BibitemShut {NoStop}%
\bibitem [{\citenamefont {Roediger}(1991)}]{RoedigerWages}%
  \BibitemOpen
  \bibfield  {author} {\bibinfo {author} {\bibfnamefont {D.~R.}\ \bibnamefont
  {Roediger}},\ }\href@noop {} {\emph {\bibinfo {title} {The Wages of
  Whiteness: Race and the Making of the American Working Class}}}\ (\bibinfo
  {publisher} {Verso},\ \bibinfo {year} {1991})\BibitemShut {NoStop}%
\bibitem [{\citenamefont {Kostas}(1998)}]{KostasLabManuals}%
  \BibitemOpen
  \bibfield  {author} {\bibinfo {author} {\bibfnamefont {N.~A.}\ \bibnamefont
  {Kostas}},\ }\emph {\bibinfo {title} {A Gender Analysis of Secondary School
  Physics Textbooks and Laboratory Manuals}},\ \href@noop {} {Ph.D. thesis},\
  \bibinfo  {school} {Lehigh University} (\bibinfo {year} {1998})\BibitemShut
  {NoStop}%
\bibitem [{\citenamefont {Brown}\ \emph {et~al.}(1998)\citenamefont {Brown},
  \citenamefont {Slater},\ and\ \citenamefont {Adams}}]{BrownBatteriesBulbs}%
  \BibitemOpen
  \bibfield  {author} {\bibinfo {author} {\bibfnamefont {T.~R.}\ \bibnamefont
  {Brown}}, \bibinfo {author} {\bibfnamefont {T.~F.}\ \bibnamefont {Slater}},\
  and\ \bibinfo {author} {\bibfnamefont {J.~P.}\ \bibnamefont {Adams}},\
  }\bibfield  {title} {\bibinfo {title} {Gender differences with batteries and
  bulbs},\ }\href@noop {} {\bibfield  {journal} {\bibinfo  {journal} {The
  Physics Teacher}\ }\textbf {\bibinfo {volume} {36}},\ \bibinfo {pages} {526}
  (\bibinfo {year} {1998})}\BibitemShut {NoStop}%
\bibitem [{\citenamefont {Wainwright}(1999)}]{WainwrightBrownResponse}%
  \BibitemOpen
  \bibfield  {author} {\bibinfo {author} {\bibfnamefont {C.~L.}\ \bibnamefont
  {Wainwright}},\ }\bibfield  {title} {\bibinfo {title} {Gender indifference},\
  }\href {https://doi.org/10.1119/1.880192} {\bibfield  {journal} {\bibinfo
  {journal} {The Physics Teacher}\ }\textbf {\bibinfo {volume} {37}},\ \bibinfo
  {pages} {131} (\bibinfo {year} {1999})}\BibitemShut {NoStop}%
\bibitem [{\citenamefont {Cotner}\ and\ \citenamefont
  {Ballen}(2017)}]{CotnerDeficit}%
  \BibitemOpen
  \bibfield  {author} {\bibinfo {author} {\bibfnamefont {S.}~\bibnamefont
  {Cotner}}\ and\ \bibinfo {author} {\bibfnamefont {C.~J.}\ \bibnamefont
  {Ballen}},\ }\bibfield  {title} {\bibinfo {title} {Can mixed assessment
  methods make biology classes more equitable?},\ }\href@noop {} {\bibfield
  {journal} {\bibinfo  {journal} {PLoS One}\ }\textbf {\bibinfo {volume}
  {12}},\ \bibinfo {pages} {e0189610} (\bibinfo {year} {2017})}\BibitemShut
  {NoStop}%
\bibitem [{\citenamefont {Exarhos}(2020)}]{ExarhosDeficitFraming}%
  \BibitemOpen
  \bibfield  {author} {\bibinfo {author} {\bibfnamefont {S.}~\bibnamefont
  {Exarhos}},\ }\bibfield  {title} {\bibinfo {title} {Anti-deficit framing of
  sociological physics education research},\ }\href@noop {} {\bibfield
  {journal} {\bibinfo  {journal} {The Physics Teacher}\ }\textbf {\bibinfo
  {volume} {58}},\ \bibinfo {pages} {461} (\bibinfo {year} {2020})}\BibitemShut
  {NoStop}%
\bibitem [{\citenamefont {Nair}\ and\ \citenamefont
  {Sawtelle}(2019)}]{NairDeficitFraming}%
  \BibitemOpen
  \bibfield  {author} {\bibinfo {author} {\bibfnamefont {A.}~\bibnamefont
  {Nair}}\ and\ \bibinfo {author} {\bibfnamefont {V.}~\bibnamefont
  {Sawtelle}},\ }\bibfield  {title} {\bibinfo {title} {Operationalizing
  relevance in physics education: Using a systems view to expand our conception
  of making physics relevant},\ }\href@noop {} {\bibfield  {journal} {\bibinfo
  {journal} {Physical Review Physics Education Research}\ }\textbf {\bibinfo
  {volume} {15}},\ \bibinfo {pages} {020121} (\bibinfo {year}
  {2019})}\BibitemShut {NoStop}%
\bibitem [{\citenamefont {Potter}\ \emph {et~al.}(2001)\citenamefont {Potter},
  \citenamefont {Leone}, \citenamefont {Ishikawa}, \citenamefont
  {Blickenstaff},\ and\ \citenamefont {Hession}}]{PotterReformedCourse}%
  \BibitemOpen
  \bibfield  {author} {\bibinfo {author} {\bibfnamefont {W.~H.}\ \bibnamefont
  {Potter}}, \bibinfo {author} {\bibfnamefont {C.~D.}\ \bibnamefont {Leone}},
  \bibinfo {author} {\bibfnamefont {C.~M.}\ \bibnamefont {Ishikawa}}, \bibinfo
  {author} {\bibfnamefont {J.}~\bibnamefont {Blickenstaff}},\ and\ \bibinfo
  {author} {\bibfnamefont {P.~L.}\ \bibnamefont {Hession}},\ }\bibfield
  {title} {\bibinfo {title} {Significant reduction in gender grade disparities
  in a reformed introductory physics course},\ }in\ \href@noop {} {\emph
  {\bibinfo {booktitle} {Physics Education Research Conference 2001}}},\
  \bibinfo {series and number} {PER Conference}\ (\bibinfo {address}
  {Rochester, New York},\ \bibinfo {year} {2001})\BibitemShut {NoStop}%
\bibitem [{\citenamefont {McKinnon}\ and\ \citenamefont
  {Potter}(2004)}]{McKinnonLabInstructions}%
  \BibitemOpen
  \bibfield  {author} {\bibinfo {author} {\bibfnamefont {M.~L.}\ \bibnamefont
  {McKinnon}}\ and\ \bibinfo {author} {\bibfnamefont {W.~H.}\ \bibnamefont
  {Potter}},\ }\bibfield  {title} {\bibinfo {title} {Preliminary results of
  gender equity variations in a large active-learning introductory physics
  course due to laboratory activity instructions},\ }in\ \href@noop {} {\emph
  {\bibinfo {booktitle} {Physics Education Research Conference 2004}}},\
  \bibinfo {series} {PER Conference}, Vol.\ \bibinfo {volume} {790}\ (\bibinfo
  {address} {Sacramento, California},\ \bibinfo {year} {2004})\ pp.\ \bibinfo
  {pages} {129--132}\BibitemShut {NoStop}%
\bibitem [{\citenamefont {Volkwyn}\ \emph {et~al.}(2008)\citenamefont
  {Volkwyn}, \citenamefont {Allie}, \citenamefont {Buffler},\ and\
  \citenamefont {Lubben}}]{WolkwynPMQ}%
  \BibitemOpen
  \bibfield  {author} {\bibinfo {author} {\bibfnamefont {T.~S.}\ \bibnamefont
  {Volkwyn}}, \bibinfo {author} {\bibfnamefont {S.}~\bibnamefont {Allie}},
  \bibinfo {author} {\bibfnamefont {A.}~\bibnamefont {Buffler}},\ and\ \bibinfo
  {author} {\bibfnamefont {F.}~\bibnamefont {Lubben}},\ }\bibfield  {title}
  {\bibinfo {title} {Impact of a conventional introductory laboratory course on
  the understanding of measurement},\ }\href
  {https://doi.org/10.1103/PhysRevSTPER.4.010108} {\bibfield  {journal}
  {\bibinfo  {journal} {Phys. Rev. ST Phys. Educ. Res.}\ }\textbf {\bibinfo
  {volume} {4}},\ \bibinfo {pages} {010108} (\bibinfo {year}
  {2008})}\BibitemShut {NoStop}%
\bibitem [{\citenamefont {Day}\ and\ \citenamefont {Bonn}(2011)}]{DayCDPA}%
  \BibitemOpen
  \bibfield  {author} {\bibinfo {author} {\bibfnamefont {J.}~\bibnamefont
  {Day}}\ and\ \bibinfo {author} {\bibfnamefont {D.}~\bibnamefont {Bonn}},\
  }\bibfield  {title} {\bibinfo {title} {Development of the concise data
  processing assessment},\ }\href
  {https://doi.org/10.1103/PhysRevSTPER.7.010114} {\bibfield  {journal}
  {\bibinfo  {journal} {Phys. Rev. ST Phys. Educ. Res.}\ }\textbf {\bibinfo
  {volume} {7}},\ \bibinfo {pages} {010114} (\bibinfo {year}
  {2011})}\BibitemShut {NoStop}%
\bibitem [{\citenamefont {Wilcox}\ and\ \citenamefont
  {Lewandowski}(2016{\natexlab{a}})}]{WilcoxECLASS}%
  \BibitemOpen
  \bibfield  {author} {\bibinfo {author} {\bibfnamefont {B.~R.}\ \bibnamefont
  {Wilcox}}\ and\ \bibinfo {author} {\bibfnamefont {H.~J.}\ \bibnamefont
  {Lewandowski}},\ }\bibfield  {title} {\bibinfo {title} {Students'
  epistemologies about experimental physics: Validating the colorado learning
  attitudes about science survey for experimental physics},\ }\href
  {https://doi.org/10.1103/PhysRevPhysEducRes.12.010123} {\bibfield  {journal}
  {\bibinfo  {journal} {Phys. Rev. Phys. Educ. Res.}\ }\textbf {\bibinfo
  {volume} {12}},\ \bibinfo {pages} {010123} (\bibinfo {year}
  {2016}{\natexlab{a}})}\BibitemShut {NoStop}%
\bibitem [{\citenamefont {Walsh}\ \emph {et~al.}(2019)\citenamefont {Walsh},
  \citenamefont {Quinn}, \citenamefont {Wieman},\ and\ \citenamefont
  {Holmes}}]{WalshPLIC}%
  \BibitemOpen
  \bibfield  {author} {\bibinfo {author} {\bibfnamefont {C.}~\bibnamefont
  {Walsh}}, \bibinfo {author} {\bibfnamefont {K.~N.}\ \bibnamefont {Quinn}},
  \bibinfo {author} {\bibfnamefont {C.}~\bibnamefont {Wieman}},\ and\ \bibinfo
  {author} {\bibfnamefont {N.~G.}\ \bibnamefont {Holmes}},\ }\bibfield  {title}
  {\bibinfo {title} {Quantifying critical thinking: Development and validation
  of the physics lab inventory of critical thinking},\ }\href
  {https://doi.org/10.1103/PhysRevPhysEducRes.15.010135} {\bibfield  {journal}
  {\bibinfo  {journal} {Phys. Rev. Phys. Educ. Res.}\ }\textbf {\bibinfo
  {volume} {15}},\ \bibinfo {pages} {010135} (\bibinfo {year}
  {2019})}\BibitemShut {NoStop}%
\bibitem [{\citenamefont {Vignal}\ \emph {et~al.}(2023)\citenamefont {Vignal},
  \citenamefont {Geschwind}, \citenamefont {Pollard}, \citenamefont
  {Henderson}, \citenamefont {Caballero},\ and\ \citenamefont
  {Lewandowski}}]{VignalSPRUCE}%
  \BibitemOpen
  \bibfield  {author} {\bibinfo {author} {\bibfnamefont {M.}~\bibnamefont
  {Vignal}}, \bibinfo {author} {\bibfnamefont {G.}~\bibnamefont {Geschwind}},
  \bibinfo {author} {\bibfnamefont {B.}~\bibnamefont {Pollard}}, \bibinfo
  {author} {\bibfnamefont {R.}~\bibnamefont {Henderson}}, \bibinfo {author}
  {\bibfnamefont {M.~D.}\ \bibnamefont {Caballero}},\ and\ \bibinfo {author}
  {\bibfnamefont {H.~J.}\ \bibnamefont {Lewandowski}},\ }\href
  {https://doi.org/10.48550/ARXIV.2302.07336} {\bibinfo {title} {Survey of
  physics reasoning on uncertainty concepts in experiments: An assessment of
  measurement uncertainty for introductory physics labs}} (\bibinfo {year}
  {2023})\BibitemShut {NoStop}%
\bibitem [{\citenamefont {Day}\ \emph {et~al.}(2016)\citenamefont {Day},
  \citenamefont {Stang}, \citenamefont {Holmes}, \citenamefont {Kumar},\ and\
  \citenamefont {Bonn}}]{DayGenderLabs}%
  \BibitemOpen
  \bibfield  {author} {\bibinfo {author} {\bibfnamefont {J.}~\bibnamefont
  {Day}}, \bibinfo {author} {\bibfnamefont {J.~B.}\ \bibnamefont {Stang}},
  \bibinfo {author} {\bibfnamefont {N.~G.}\ \bibnamefont {Holmes}}, \bibinfo
  {author} {\bibfnamefont {D.}~\bibnamefont {Kumar}},\ and\ \bibinfo {author}
  {\bibfnamefont {D.~A.}\ \bibnamefont {Bonn}},\ }\bibfield  {title} {\bibinfo
  {title} {Gender gaps and gendered action in a first-year physics
  laboratory},\ }\href {https://doi.org/10.1103/PhysRevPhysEducRes.12.020104}
  {\bibfield  {journal} {\bibinfo  {journal} {Phys. Rev. Phys. Educ. Res.}\
  }\textbf {\bibinfo {volume} {12}},\ \bibinfo {pages} {020104} (\bibinfo
  {year} {2016})}\BibitemShut {NoStop}%
\bibitem [{\citenamefont {Wilcox}\ and\ \citenamefont
  {Lewandowski}(2016{\natexlab{b}})}]{WilcoxECLASSgender}%
  \BibitemOpen
  \bibfield  {author} {\bibinfo {author} {\bibfnamefont {B.~R.}\ \bibnamefont
  {Wilcox}}\ and\ \bibinfo {author} {\bibfnamefont {H.}~\bibnamefont
  {Lewandowski}},\ }\bibfield  {title} {\bibinfo {title} {Research-based
  assessment of students' beliefs about experimental physics: When is gender a
  factor?},\ }\href@noop {} {\bibfield  {journal} {\bibinfo  {journal} {Phys.
  Rev. Phys. Educ. Res.}\ }\textbf {\bibinfo {volume} {12}},\ \bibinfo {pages}
  {020130} (\bibinfo {year} {2016}{\natexlab{b}})}\BibitemShut {NoStop}%
\bibitem [{\citenamefont {Doucette}\ \emph {et~al.}(2022)\citenamefont
  {Doucette}, \citenamefont {Clark},\ and\ \citenamefont
  {Singh}}]{DoucetteCJP}%
  \BibitemOpen
  \bibfield  {author} {\bibinfo {author} {\bibfnamefont {D.}~\bibnamefont
  {Doucette}}, \bibinfo {author} {\bibfnamefont {R.}~\bibnamefont {Clark}},\
  and\ \bibinfo {author} {\bibfnamefont {C.}~\bibnamefont {Singh}},\ }\bibfield
   {title} {\bibinfo {title} {Students’ attitudes toward experimental physics
  in a conceptual inquiry-based introductory physics lab},\ }\href@noop {}
  {\bibfield  {journal} {\bibinfo  {journal} {Canadian Journal of Physics}\
  }\textbf {\bibinfo {volume} {100}},\ \bibinfo {pages} {292} (\bibinfo {year}
  {2022})}\BibitemShut {NoStop}%
\bibitem [{\citenamefont {Wilcox}\ and\ \citenamefont
  {Lewandowski}(2017)}]{WilcoxECLASSSkills}%
  \BibitemOpen
  \bibfield  {author} {\bibinfo {author} {\bibfnamefont {B.~R.}\ \bibnamefont
  {Wilcox}}\ and\ \bibinfo {author} {\bibfnamefont {H.~J.}\ \bibnamefont
  {Lewandowski}},\ }\bibfield  {title} {\bibinfo {title} {Developing skills
  versus reinforcing concepts in physics labs: Insight from a survey of
  students' beliefs about experimental physics},\ }\href
  {https://doi.org/10.1103/PhysRevPhysEducRes.13.010108} {\bibfield  {journal}
  {\bibinfo  {journal} {Phys. Rev. Phys. Educ. Res.}\ }\textbf {\bibinfo
  {volume} {13}},\ \bibinfo {pages} {010108} (\bibinfo {year}
  {2017})}\BibitemShut {NoStop}%
\bibitem [{\citenamefont {Walsh}\ \emph {et~al.}(2022)\citenamefont {Walsh},
  \citenamefont {Lewandowski},\ and\ \citenamefont {Holmes}}]{WalshSkillsLabs}%
  \BibitemOpen
  \bibfield  {author} {\bibinfo {author} {\bibfnamefont {C.}~\bibnamefont
  {Walsh}}, \bibinfo {author} {\bibfnamefont {H.~J.}\ \bibnamefont
  {Lewandowski}},\ and\ \bibinfo {author} {\bibfnamefont {N.~G.}\ \bibnamefont
  {Holmes}},\ }\bibfield  {title} {\bibinfo {title} {Skills-focused lab
  instruction improves critical thinking skills and experimentation views for
  all students},\ }\href {https://doi.org/10.1103/PhysRevPhysEducRes.18.010128}
  {\bibfield  {journal} {\bibinfo  {journal} {Phys. Rev. Phys. Educ. Res.}\
  }\textbf {\bibinfo {volume} {18}},\ \bibinfo {pages} {010128} (\bibinfo
  {year} {2022})}\BibitemShut {NoStop}%
\bibitem [{\citenamefont {Sulaiman}\ \emph {et~al.}(2023)\citenamefont
  {Sulaiman}, \citenamefont {Werth},\ and\ \citenamefont
  {Lewandowski}}]{SulaimanECLASSgender2}%
  \BibitemOpen
  \bibfield  {author} {\bibinfo {author} {\bibfnamefont {N.}~\bibnamefont
  {Sulaiman}}, \bibinfo {author} {\bibfnamefont {A.}~\bibnamefont {Werth}},\
  and\ \bibinfo {author} {\bibfnamefont {H.~J.}\ \bibnamefont {Lewandowski}},\
  }\bibfield  {title} {\bibinfo {title} {Students' views about experimental
  physics in a large-enrollment introductory lab focused on experimental
  scientific practices},\ }\href
  {https://doi.org/10.1103/PhysRevPhysEducRes.19.010116} {\bibfield  {journal}
  {\bibinfo  {journal} {Phys. Rev. Phys. Educ. Res.}\ }\textbf {\bibinfo
  {volume} {19}},\ \bibinfo {pages} {010116} (\bibinfo {year}
  {2023})}\BibitemShut {NoStop}%
\bibitem [{\citenamefont {Kohl}\ and\ \citenamefont {Kuo}(2009)}]{KohlStudio}%
  \BibitemOpen
  \bibfield  {author} {\bibinfo {author} {\bibfnamefont {P.~B.}\ \bibnamefont
  {Kohl}}\ and\ \bibinfo {author} {\bibfnamefont {V.}~\bibnamefont {Kuo}},\
  }\bibfield  {title} {\bibinfo {title} {Introductory physics gender gaps: Pre-
  and post-studio transition},\ }in\ \href@noop {} {\emph {\bibinfo {booktitle}
  {Physics Education Research Conference 2009}}},\ \bibinfo {series} {PER
  Conference}, Vol.\ \bibinfo {volume} {1179}\ (\bibinfo {address} {Ann Arbor,
  Michigan},\ \bibinfo {year} {2009})\ pp.\ \bibinfo {pages}
  {173--176}\BibitemShut {NoStop}%
\bibitem [{\citenamefont {Laws}\ \emph {et~al.}(1999)\citenamefont {Laws},
  \citenamefont {Rosborough},\ and\ \citenamefont
  {Poodry}}]{LawsWorkshopPhysicsWomensResponses}%
  \BibitemOpen
  \bibfield  {author} {\bibinfo {author} {\bibfnamefont {P.~W.}\ \bibnamefont
  {Laws}}, \bibinfo {author} {\bibfnamefont {P.~J.}\ \bibnamefont
  {Rosborough}},\ and\ \bibinfo {author} {\bibfnamefont {F.~J.}\ \bibnamefont
  {Poodry}},\ }\bibfield  {title} {\bibinfo {title} {Women's responses to an
  activity-based introductory physics program},\ }\href@noop {} {\bibfield
  {journal} {\bibinfo  {journal} {Am. J. Phys.}\ }\textbf {\bibinfo {volume}
  {67}},\ \bibinfo {pages} {32} (\bibinfo {year} {1999})}\BibitemShut {NoStop}%
\bibitem [{\citenamefont {Beichner}(2007)}]{BeichnerSCALEUP}%
  \BibitemOpen
  \bibfield  {author} {\bibinfo {author} {\bibfnamefont {R.}~\bibnamefont
  {Beichner}},\ }\bibfield  {title} {\bibinfo {title} {The student-centered
  activities for large enrollment undergraduate programs (scale-up) project},\
  }in\ \href@noop {} {\emph {\bibinfo {booktitle} {Research-Based Reform of
  University Physics}}},\ Vol.~\bibinfo {volume} {1},\ \bibinfo {editor}
  {edited by\ \bibinfo {editor} {\bibfnamefont {E.~F.}\ \bibnamefont {Redish}}\
  and\ \bibinfo {editor} {\bibfnamefont {P.}~\bibnamefont {Cooney}}}\ (\bibinfo
   {publisher} {American Association of Physics Teachers},\ \bibinfo {address}
  {College Park},\ \bibinfo {year} {2007})\BibitemShut {NoStop}%
\bibitem [{\citenamefont {Brahmia}(2008)}]{BrahmiaImprovingLearning}%
  \BibitemOpen
  \bibfield  {author} {\bibinfo {author} {\bibfnamefont {S.~W.}\ \bibnamefont
  {Brahmia}},\ }\bibfield  {title} {\bibinfo {title} {Improving learning for
  underrepresented groups in physics for engineering majors},\ }\href
  {https://doi.org/10.1063/1.3021279} {\bibfield  {journal} {\bibinfo
  {journal} {AIP Conference Proceedings}\ }\textbf {\bibinfo {volume} {1064}},\
  \bibinfo {pages} {7} (\bibinfo {year} {2008})}\BibitemShut {NoStop}%
\bibitem [{\citenamefont {Traxler}\ and\ \citenamefont
  {Brewe}(2015)}]{TraxlerModeling}%
  \BibitemOpen
  \bibfield  {author} {\bibinfo {author} {\bibfnamefont {A.}~\bibnamefont
  {Traxler}}\ and\ \bibinfo {author} {\bibfnamefont {E.}~\bibnamefont
  {Brewe}},\ }\bibfield  {title} {\bibinfo {title} {Equity investigation of
  attitudinal shifts in introductory physics},\ }\href
  {https://doi.org/10.1103/PhysRevSTPER.11.020132} {\bibfield  {journal}
  {\bibinfo  {journal} {Phys. Rev. ST Phys. Educ. Res.}\ }\textbf {\bibinfo
  {volume} {11}},\ \bibinfo {pages} {020132} (\bibinfo {year}
  {2015})}\BibitemShut {NoStop}%
\bibitem [{\citenamefont {Rodriguez}\ \emph {et~al.}(2016)\citenamefont
  {Rodriguez}, \citenamefont {Potvin},\ and\ \citenamefont
  {Kramer}}]{RodriguezGenderReformed}%
  \BibitemOpen
  \bibfield  {author} {\bibinfo {author} {\bibfnamefont {I.}~\bibnamefont
  {Rodriguez}}, \bibinfo {author} {\bibfnamefont {G.}~\bibnamefont {Potvin}},\
  and\ \bibinfo {author} {\bibfnamefont {L.~H.}\ \bibnamefont {Kramer}},\
  }\bibfield  {title} {\bibinfo {title} {How gender and reformed introductory
  physics impacts student success in advanced physics courses and continuation
  in the physics major},\ }\href
  {https://doi.org/10.1103/PhysRevPhysEducRes.12.020118} {\bibfield  {journal}
  {\bibinfo  {journal} {Phys. Rev. Phys. Educ. Res.}\ }\textbf {\bibinfo
  {volume} {12}},\ \bibinfo {pages} {020118} (\bibinfo {year}
  {2016})}\BibitemShut {NoStop}%
\bibitem [{\citenamefont {Brewe}\ \emph {et~al.}(2010)\citenamefont {Brewe},
  \citenamefont {Sawtelle}, \citenamefont {Kramer}, \citenamefont {O'Brien},
  \citenamefont {Rodriguez},\ and\ \citenamefont
  {Pamel\'a}}]{BreweTowardEquity}%
  \BibitemOpen
  \bibfield  {author} {\bibinfo {author} {\bibfnamefont {E.}~\bibnamefont
  {Brewe}}, \bibinfo {author} {\bibfnamefont {V.}~\bibnamefont {Sawtelle}},
  \bibinfo {author} {\bibfnamefont {L.~H.}\ \bibnamefont {Kramer}}, \bibinfo
  {author} {\bibfnamefont {G.~E.}\ \bibnamefont {O'Brien}}, \bibinfo {author}
  {\bibfnamefont {I.}~\bibnamefont {Rodriguez}},\ and\ \bibinfo {author}
  {\bibfnamefont {P.}~\bibnamefont {Pamel\'a}},\ }\bibfield  {title} {\bibinfo
  {title} {Toward equity through participation in modeling instruction in
  introductory university physics},\ }\href
  {https://doi.org/10.1103/PhysRevSTPER.6.010106} {\bibfield  {journal}
  {\bibinfo  {journal} {Phys. Rev. ST Phys. Educ. Res.}\ }\textbf {\bibinfo
  {volume} {6}},\ \bibinfo {pages} {010106} (\bibinfo {year}
  {2010})}\BibitemShut {NoStop}%
\bibitem [{\citenamefont {McPadden}\ and\ \citenamefont
  {Brewe}(2014)}]{McPaddenMIGender}%
  \BibitemOpen
  \bibfield  {author} {\bibinfo {author} {\bibfnamefont {D.~R.}\ \bibnamefont
  {McPadden}}\ and\ \bibinfo {author} {\bibfnamefont {E.}~\bibnamefont
  {Brewe}},\ }\bibfield  {title} {\bibinfo {title} {The impacts of instructor
  and student gender on student performance in introductory modeling
  instruction courses},\ }in\ \href@noop {} {\emph {\bibinfo {booktitle}
  {Physics Education Research Conference 2014}}},\ \bibinfo {series and number}
  {PER Conference}\ (\bibinfo {address} {Minneapolis, MN},\ \bibinfo {year}
  {2014})\ pp.\ \bibinfo {pages} {183--186}\BibitemShut {NoStop}%
\bibitem [{\citenamefont {Kepple}(2017)}]{KeppleDistance}%
  \BibitemOpen
  \bibfield  {author} {\bibinfo {author} {\bibfnamefont {C.~M.~J.}\
  \bibnamefont {Kepple}},\ }\emph {\bibinfo {title} {Distance Learning in
  Physics: An Investigation of Possible Benefits to Female Students in an
  Online Laboratory Environment}},\ \href@noop {} {Master's thesis},\ \bibinfo
  {school} {Portland State University} (\bibinfo {year} {2017})\BibitemShut
  {NoStop}%
\bibitem [{\citenamefont {Rosen}\ and\ \citenamefont
  {Kelly}(2020)}]{RosenEpistemology}%
  \BibitemOpen
  \bibfield  {author} {\bibinfo {author} {\bibfnamefont {D.~J.}\ \bibnamefont
  {Rosen}}\ and\ \bibinfo {author} {\bibfnamefont {A.~M.}\ \bibnamefont
  {Kelly}},\ }\bibfield  {title} {\bibinfo {title} {Epistemology,
  socialization, help seeking, and gender-based views in in-person and online,
  hands-on undergraduate physics laboratories},\ }\href
  {https://doi.org/10.1103/PhysRevPhysEducRes.16.020116} {\bibfield  {journal}
  {\bibinfo  {journal} {Phys. Rev. Phys. Educ. Res.}\ }\textbf {\bibinfo
  {volume} {16}},\ \bibinfo {pages} {020116} (\bibinfo {year}
  {2020})}\BibitemShut {NoStop}%
\bibitem [{\citenamefont {Radulovi{\'c}}\ \emph {et~al.}(2022)\citenamefont
  {Radulovi{\'c}}, \citenamefont {{\v{Z}}upanec}, \citenamefont
  {Stojanovi{\'c}},\ and\ \citenamefont
  {Budi{\'c}}}]{RadulovicVirtualMotivation}%
  \BibitemOpen
  \bibfield  {author} {\bibinfo {author} {\bibfnamefont {B.}~\bibnamefont
  {Radulovi{\'c}}}, \bibinfo {author} {\bibfnamefont {V.}~\bibnamefont
  {{\v{Z}}upanec}}, \bibinfo {author} {\bibfnamefont {M.}~\bibnamefont
  {Stojanovi{\'c}}},\ and\ \bibinfo {author} {\bibfnamefont {S.}~\bibnamefont
  {Budi{\'c}}},\ }\bibfield  {title} {\bibinfo {title} {Gender motivational gap
  and contribution of different teaching approaches to female students’
  motivation to learn physics},\ }\href@noop {} {\bibfield  {journal} {\bibinfo
   {journal} {Scientific Reports}\ }\textbf {\bibinfo {volume} {12}},\ \bibinfo
  {pages} {18224} (\bibinfo {year} {2022})}\BibitemShut {NoStop}%
\bibitem [{\citenamefont {Burkam}\ \emph {et~al.}(1997)\citenamefont {Burkam},
  \citenamefont {Lee},\ and\ \citenamefont {Smerdon}}]{BurkamHighSchool}%
  \BibitemOpen
  \bibfield  {author} {\bibinfo {author} {\bibfnamefont {D.~T.}\ \bibnamefont
  {Burkam}}, \bibinfo {author} {\bibfnamefont {V.~E.}\ \bibnamefont {Lee}},\
  and\ \bibinfo {author} {\bibfnamefont {B.~A.}\ \bibnamefont {Smerdon}},\
  }\bibfield  {title} {\bibinfo {title} {Gender and science learning early in
  high school: Subject matter and laboratory experiences},\ }\href@noop {}
  {\bibfield  {journal} {\bibinfo  {journal} {American Educational Research
  Journal}\ }\textbf {\bibinfo {volume} {34}},\ \bibinfo {pages} {297}
  (\bibinfo {year} {1997})}\BibitemShut {NoStop}%
\bibitem [{\citenamefont {Hazari}\ \emph {et~al.}(2008)\citenamefont {Hazari},
  \citenamefont {Sadler},\ and\ \citenamefont {Tai}}]{HazariGenderDiffTPT}%
  \BibitemOpen
  \bibfield  {author} {\bibinfo {author} {\bibfnamefont {Z.}~\bibnamefont
  {Hazari}}, \bibinfo {author} {\bibfnamefont {P.~M.}\ \bibnamefont {Sadler}},\
  and\ \bibinfo {author} {\bibfnamefont {R.~H.}\ \bibnamefont {Tai}},\
  }\bibfield  {title} {\bibinfo {title} {Gender differences in the high school
  and affective experiences of introductory college physics students},\
  }\href@noop {} {\bibfield  {journal} {\bibinfo  {journal} {The Physics
  Teacher}\ }\textbf {\bibinfo {volume} {46}},\ \bibinfo {pages} {423}
  (\bibinfo {year} {2008})}\BibitemShut {NoStop}%
\bibitem [{\citenamefont {Stoeckel}(2020)}]{StoeckelConfidence}%
  \BibitemOpen
  \bibfield  {author} {\bibinfo {author} {\bibfnamefont {M.~R.}\ \bibnamefont
  {Stoeckel}},\ }\bibfield  {title} {\bibinfo {title} {Gender, self-assessment,
  and classroom experiences in {AP Physics 1}},\ }\href@noop {} {\bibfield
  {journal} {\bibinfo  {journal} {The Physics Teacher}\ }\textbf {\bibinfo
  {volume} {58}},\ \bibinfo {pages} {399} (\bibinfo {year} {2020})}\BibitemShut
  {NoStop}%
\bibitem [{\citenamefont {Thomas}(2023)}]{ThomasPhD}%
  \BibitemOpen
  \bibfield  {author} {\bibinfo {author} {\bibfnamefont {D.}~\bibnamefont
  {Thomas}},\ }\emph {\bibinfo {title} {Exploring the Contextually Situated
  Experiences, Perceptions, Beliefs, and Intentions of College Physics Majors
  in Physics Courses that Incorporate an Interactive Instructional
  Approach}},\ \href@noop {} {Ph.D. thesis},\ \bibinfo  {school} {North
  Carolina State University} (\bibinfo {year} {2023})\BibitemShut {NoStop}%
\bibitem [{\citenamefont {Gunter}\ \emph {et~al.}(2010)\citenamefont {Gunter},
  \citenamefont {Spiczak},\ and\ \citenamefont {Madsen}}]{GunterAstroLab}%
  \BibitemOpen
  \bibfield  {author} {\bibinfo {author} {\bibfnamefont {R.}~\bibnamefont
  {Gunter}}, \bibinfo {author} {\bibfnamefont {G.}~\bibnamefont {Spiczak}},\
  and\ \bibinfo {author} {\bibfnamefont {J.}~\bibnamefont {Madsen}},\
  }\bibfield  {title} {\bibinfo {title} {Cosmic collaboration in an
  undergraduate astrophysics laboratory},\ }\href@noop {} {\bibfield  {journal}
  {\bibinfo  {journal} {American Journal of Physics}\ }\textbf {\bibinfo
  {volume} {78}},\ \bibinfo {pages} {1035} (\bibinfo {year}
  {2010})}\BibitemShut {NoStop}%
\bibitem [{\citenamefont {Traweek}(1988)}]{TraweekBeamtimes}%
  \BibitemOpen
  \bibfield  {author} {\bibinfo {author} {\bibfnamefont {S.}~\bibnamefont
  {Traweek}},\ }\href@noop {} {\emph {\bibinfo {title} {Beamtimes and
  Lifetimes}}}\ (\bibinfo  {publisher} {Harvard University Press},\ \bibinfo
  {year} {1988})\BibitemShut {NoStop}%
\bibitem [{\citenamefont {Pettersson}(2011)}]{PetterssonMaking}%
  \BibitemOpen
  \bibfield  {author} {\bibinfo {author} {\bibfnamefont {H.}~\bibnamefont
  {Pettersson}},\ }\bibfield  {title} {\bibinfo {title} {Making masculinity in
  plasma physics: Machines, labour and experiments},\ }\href@noop {} {\bibfield
   {journal} {\bibinfo  {journal} {Science \& Technology Studies}\ }\textbf
  {\bibinfo {volume} {24}},\ \bibinfo {pages} {47} (\bibinfo {year}
  {2011})}\BibitemShut {NoStop}%
\bibitem [{\citenamefont {Danielsson}(2009)}]{DanielssonDoing}%
  \BibitemOpen
  \bibfield  {author} {\bibinfo {author} {\bibfnamefont {A.~T.}\ \bibnamefont
  {Danielsson}},\ }\emph {\bibinfo {title} {Doing Physics--Doing Gender: An
  Exploration of Physics Students' Identity Constitution in the Context of
  Laboratory Work}},\ \href@noop {} {Ph.D. thesis},\ \bibinfo  {school}
  {Uppsala University} (\bibinfo {year} {2009})\BibitemShut {NoStop}%
\bibitem [{\citenamefont {Danielsson}(2012)}]{DanielssonExploring}%
  \BibitemOpen
  \bibfield  {author} {\bibinfo {author} {\bibfnamefont {A.~T.}\ \bibnamefont
  {Danielsson}},\ }\bibfield  {title} {\bibinfo {title} {Exploring woman
  university physics students ‘doing gender’and ‘doing physics’},\
  }\href@noop {} {\bibfield  {journal} {\bibinfo  {journal} {Gender and
  Education}\ }\textbf {\bibinfo {volume} {24}},\ \bibinfo {pages} {25}
  (\bibinfo {year} {2012})}\BibitemShut {NoStop}%
\bibitem [{\citenamefont {Kalender}\ \emph {et~al.}(2021)\citenamefont
  {Kalender}, \citenamefont {Stump}, \citenamefont {Hubenig},\ and\
  \citenamefont {Holmes}}]{KalenderAgencyLabs}%
  \BibitemOpen
  \bibfield  {author} {\bibinfo {author} {\bibfnamefont {Z.~Y.}\ \bibnamefont
  {Kalender}}, \bibinfo {author} {\bibfnamefont {E.}~\bibnamefont {Stump}},
  \bibinfo {author} {\bibfnamefont {K.}~\bibnamefont {Hubenig}},\ and\ \bibinfo
  {author} {\bibfnamefont {N.~G.}\ \bibnamefont {Holmes}},\ }\bibfield  {title}
  {\bibinfo {title} {Restructuring physics labs to cultivate sense of student
  agency},\ }\href {https://doi.org/10.1103/PhysRevPhysEducRes.17.020128}
  {\bibfield  {journal} {\bibinfo  {journal} {Phys. Rev. Phys. Educ. Res.}\
  }\textbf {\bibinfo {volume} {17}},\ \bibinfo {pages} {020128} (\bibinfo
  {year} {2021})}\BibitemShut {NoStop}%
\bibitem [{\citenamefont {Kalender}\ \emph {et~al.}(2020)\citenamefont
  {Kalender}, \citenamefont {Stein},\ and\ \citenamefont
  {Holmes}}]{KalenderAgencyPERC}%
  \BibitemOpen
  \bibfield  {author} {\bibinfo {author} {\bibfnamefont {Z.~Y.}\ \bibnamefont
  {Kalender}}, \bibinfo {author} {\bibfnamefont {M.~M.}\ \bibnamefont
  {Stein}},\ and\ \bibinfo {author} {\bibfnamefont {N.~G.}\ \bibnamefont
  {Holmes}},\ }\bibfield  {title} {\bibinfo {title} {Sense of agency, gender,
  and students’ perception in open-ended physics labs},\ }in\ \href@noop {}
  {\emph {\bibinfo {booktitle} {Physics Education Research Conference 2020}}},\
  \bibinfo {series and number} {PER Conference}\ (\bibinfo {address} {Virtual
  Conference},\ \bibinfo {year} {2020})\ pp.\ \bibinfo {pages}
  {259--265}\BibitemShut {NoStop}%
\bibitem [{\citenamefont {Kinnischtzke}\ and\ \citenamefont
  {Smith}(2021)}]{KinnischtzkeAgencySE}%
  \BibitemOpen
  \bibfield  {author} {\bibinfo {author} {\bibfnamefont {M.}~\bibnamefont
  {Kinnischtzke}}\ and\ \bibinfo {author} {\bibfnamefont {E.~M.}\ \bibnamefont
  {Smith}},\ }\bibfield  {title} {\bibinfo {title} {Investigating relationships
  between emotional states and self-efficacy, agency, and interest in
  introductory labs},\ }in\ \href@noop {} {\emph {\bibinfo {booktitle} {Physics
  Education Research Conference 2021}}},\ \bibinfo {series and number} {PER
  Conference}\ (\bibinfo {address} {Virtual Conference},\ \bibinfo {year}
  {2021})\ pp.\ \bibinfo {pages} {209--214}\BibitemShut {NoStop}%
\bibitem [{\citenamefont {Stump}\ and\ \citenamefont
  {Holmes}(2022)}]{StumpRecog}%
  \BibitemOpen
  \bibfield  {author} {\bibinfo {author} {\bibfnamefont {E.~M.}\ \bibnamefont
  {Stump}}\ and\ \bibinfo {author} {\bibfnamefont {N.~G.}\ \bibnamefont
  {Holmes}},\ }\bibfield  {title} {\bibinfo {title} {Student views of what
  counts as doing physics in the lab},\ }in\ \href@noop {} {\emph {\bibinfo
  {booktitle} {Physics Education Research Conference 2022}}},\ \bibinfo {series
  and number} {PER Conference}\ (\bibinfo {address} {Grand Rapids, MI},\
  \bibinfo {year} {2022})\ pp.\ \bibinfo {pages} {444--450}\BibitemShut
  {NoStop}%
\bibitem [{\citenamefont {Doucette}\ and\ \citenamefont
  {Singh}(2020)}]{DoucetteReflectionsTwoWomen}%
  \BibitemOpen
  \bibfield  {author} {\bibinfo {author} {\bibfnamefont {D.}~\bibnamefont
  {Doucette}}\ and\ \bibinfo {author} {\bibfnamefont {C.}~\bibnamefont
  {Singh}},\ }\bibfield  {title} {\bibinfo {title} {Why are there so few women
  in physics? {R}eflections on the experiences of two women},\ }\href
  {https://doi.org/10.1119/1.5145518} {\bibfield  {journal} {\bibinfo
  {journal} {The Physics Teacher}\ }\textbf {\bibinfo {volume} {58}},\ \bibinfo
  {pages} {297} (\bibinfo {year} {2020})}\BibitemShut {NoStop}%
\bibitem [{\citenamefont {Li}\ and\ \citenamefont {Singh}(2023)}]{LiImpact}%
  \BibitemOpen
  \bibfield  {author} {\bibinfo {author} {\bibfnamefont {Y.}~\bibnamefont
  {Li}}\ and\ \bibinfo {author} {\bibfnamefont {C.}~\bibnamefont {Singh}},\
  }\bibfield  {title} {\bibinfo {title} {The impact of perceived recognition by
  physics instructors on women's self-efficacy and interest},\ }\href@noop {}
  {\bibfield  {journal} {\bibinfo  {journal} {arXiv preprint arXiv:2303.07239}\
  } (\bibinfo {year} {2023})}\BibitemShut {NoStop}%
\bibitem [{\citenamefont {Gunawan}\ \emph {et~al.}(2018)\citenamefont
  {Gunawan}, \citenamefont {Suranti}, \citenamefont {Nisrina}, \citenamefont
  {Herayanti},\ and\ \citenamefont {Rahmatiah}}]{GunawanCreativity}%
  \BibitemOpen
  \bibfield  {author} {\bibinfo {author} {\bibfnamefont {G.}~\bibnamefont
  {Gunawan}}, \bibinfo {author} {\bibfnamefont {N.}~\bibnamefont {Suranti}},
  \bibinfo {author} {\bibfnamefont {N.}~\bibnamefont {Nisrina}}, \bibinfo
  {author} {\bibfnamefont {L.}~\bibnamefont {Herayanti}},\ and\ \bibinfo
  {author} {\bibfnamefont {R.}~\bibnamefont {Rahmatiah}},\ }\bibfield  {title}
  {\bibinfo {title} {The effect of virtual lab and gender toward students’
  creativity of physics in senior high school},\ }in\ \href@noop {} {\emph
  {\bibinfo {booktitle} {Journal of Physics: Conference Series}}},\ Vol.\
  \bibinfo {volume} {1108}\ (\bibinfo {organization} {IOP Publishing},\
  \bibinfo {year} {2018})\BibitemShut {NoStop}%
\bibitem [{\citenamefont {Funkhouser}(2019)}]{FunkhouserPhD}%
  \BibitemOpen
  \bibfield  {author} {\bibinfo {author} {\bibfnamefont {K.~M.}\ \bibnamefont
  {Funkhouser}},\ }\href@noop {} {\emph {\bibinfo {title} {Examining Physics
  Identity in Laboratory Settings through Survey Development}}}\ (\bibinfo
  {publisher} {Michigan State University},\ \bibinfo {year} {2019})\BibitemShut
  {NoStop}%
\bibitem [{\citenamefont {Lassen}\ \emph {et~al.}(2021)\citenamefont {Lassen},
  \citenamefont {Arielle-Evans}, \citenamefont {Ríos}, \citenamefont
  {Lewandowski},\ and\ \citenamefont {Dounas-Frazer}}]{LassenOwnership}%
  \BibitemOpen
  \bibfield  {author} {\bibinfo {author} {\bibfnamefont {I.~C.}\ \bibnamefont
  {Lassen}}, \bibinfo {author} {\bibfnamefont {A.}~\bibnamefont
  {Arielle-Evans}}, \bibinfo {author} {\bibfnamefont {L.}~\bibnamefont
  {Ríos}}, \bibinfo {author} {\bibfnamefont {H.~J.}\ \bibnamefont
  {Lewandowski}},\ and\ \bibinfo {author} {\bibfnamefont {D.}~\bibnamefont
  {Dounas-Frazer}},\ }\bibfield  {title} {\bibinfo {title} {Student ownership
  and understanding of multi-week final projects},\ }in\ \href@noop {} {\emph
  {\bibinfo {booktitle} {Physics Education Research Conference 2021}}},\
  \bibinfo {series and number} {PER Conference}\ (\bibinfo {address} {Virtual
  Conference},\ \bibinfo {year} {2021})\ pp.\ \bibinfo {pages}
  {221--226}\BibitemShut {NoStop}%
\bibitem [{\citenamefont {Jeon}\ \emph {et~al.}(2020)\citenamefont {Jeon},
  \citenamefont {Kalender}, \citenamefont {Sayre},\ and\ \citenamefont
  {Holmes}}]{JeonInchargeness}%
  \BibitemOpen
  \bibfield  {author} {\bibinfo {author} {\bibfnamefont {S.~M.}\ \bibnamefont
  {Jeon}}, \bibinfo {author} {\bibfnamefont {Z.~Y.}\ \bibnamefont {Kalender}},
  \bibinfo {author} {\bibfnamefont {E.}~\bibnamefont {Sayre}},\ and\ \bibinfo
  {author} {\bibfnamefont {N.~G.}\ \bibnamefont {Holmes}},\ }\bibfield  {title}
  {\bibinfo {title} {How do gender and inchargeness interact to affect equity
  in lab group interactions?},\ }in\ \href@noop {} {\emph {\bibinfo {booktitle}
  {Physics Education Research Conference 2020}}},\ \bibinfo {series and number}
  {PER Conference}\ (\bibinfo {address} {Virtual Conference},\ \bibinfo {year}
  {2020})\ pp.\ \bibinfo {pages} {240--245}\BibitemShut {NoStop}%
\bibitem [{\citenamefont {Sundstrom}\ \emph
  {et~al.}(2022{\natexlab{a}})\citenamefont {Sundstrom}, \citenamefont
  {Schang}, \citenamefont {Heim},\ and\ \citenamefont
  {Holmes}}]{SundstromNetworkFormation}%
  \BibitemOpen
  \bibfield  {author} {\bibinfo {author} {\bibfnamefont {M.}~\bibnamefont
  {Sundstrom}}, \bibinfo {author} {\bibfnamefont {A.}~\bibnamefont {Schang}},
  \bibinfo {author} {\bibfnamefont {A.~B.}\ \bibnamefont {Heim}},\ and\
  \bibinfo {author} {\bibfnamefont {N.~G.}\ \bibnamefont {Holmes}},\ }\bibfield
   {title} {\bibinfo {title} {Understanding interaction network formation
  across instructional contexts in remote physics courses},\ }\href
  {https://doi.org/10.1103/PhysRevPhysEducRes.18.020141} {\bibfield  {journal}
  {\bibinfo  {journal} {Phys. Rev. Phys. Educ. Res.}\ }\textbf {\bibinfo
  {volume} {18}},\ \bibinfo {pages} {020141} (\bibinfo {year}
  {2022}{\natexlab{a}})}\BibitemShut {NoStop}%
\bibitem [{\citenamefont {Sundstrom}\ \emph
  {et~al.}(2022{\natexlab{b}})\citenamefont {Sundstrom}, \citenamefont {Wu},
  \citenamefont {Walsh}, \citenamefont {Heim},\ and\ \citenamefont
  {Holmes}}]{SundstromIntergroupNetworks}%
  \BibitemOpen
  \bibfield  {author} {\bibinfo {author} {\bibfnamefont {M.}~\bibnamefont
  {Sundstrom}}, \bibinfo {author} {\bibfnamefont {D.~G.}\ \bibnamefont {Wu}},
  \bibinfo {author} {\bibfnamefont {C.}~\bibnamefont {Walsh}}, \bibinfo
  {author} {\bibfnamefont {A.~B.}\ \bibnamefont {Heim}},\ and\ \bibinfo
  {author} {\bibfnamefont {N.~G.}\ \bibnamefont {Holmes}},\ }\bibfield  {title}
  {\bibinfo {title} {Examining the effects of lab instruction and gender
  composition on intergroup interaction networks in introductory physics
  labs},\ }\href {https://doi.org/10.1103/PhysRevPhysEducRes.18.010102}
  {\bibfield  {journal} {\bibinfo  {journal} {Phys. Rev. Phys. Educ. Res.}\
  }\textbf {\bibinfo {volume} {18}},\ \bibinfo {pages} {010102} (\bibinfo
  {year} {2022}{\natexlab{b}})}\BibitemShut {NoStop}%
\bibitem [{\citenamefont {Doucette}\ and\ \citenamefont
  {Singh}(2023)}]{DoucetteGroupwork}%
  \BibitemOpen
  \bibfield  {author} {\bibinfo {author} {\bibfnamefont {D.}~\bibnamefont
  {Doucette}}\ and\ \bibinfo {author} {\bibfnamefont {C.}~\bibnamefont
  {Singh}},\ }\bibfield  {title} {\bibinfo {title} {Making lab group work
  equitable and inclusive},\ }\href@noop {} {\bibfield  {journal} {\bibinfo
  {journal} {Journal of College Science Teaching}\ }\textbf {\bibinfo {volume}
  {52}} (\bibinfo {year} {2023})}\BibitemShut {NoStop}%
\bibitem [{\citenamefont {Holmes}\ \emph {et~al.}(2014)\citenamefont {Holmes},
  \citenamefont {Roll},\ and\ \citenamefont {Bonn}}]{HolmesGenderLabs}%
  \BibitemOpen
  \bibfield  {author} {\bibinfo {author} {\bibfnamefont {N.}~\bibnamefont
  {Holmes}}, \bibinfo {author} {\bibfnamefont {I.}~\bibnamefont {Roll}},\ and\
  \bibinfo {author} {\bibfnamefont {D.}~\bibnamefont {Bonn}},\ }\bibfield
  {title} {\bibinfo {title} {Participating in the physics lab: Does gender
  matter?},\ }\href@noop {} {\bibfield  {journal} {\bibinfo  {journal} {Physics
  in Canada, Special Issue on Physics Education Research}\ }\textbf {\bibinfo
  {volume} {70}},\ \bibinfo {pages} {84} (\bibinfo {year} {2014})}\BibitemShut
  {NoStop}%
\bibitem [{\citenamefont {Quinn}\ \emph {et~al.}(2020)\citenamefont {Quinn},
  \citenamefont {Kelley}, \citenamefont {McGill}, \citenamefont {Smith},
  \citenamefont {Whipps},\ and\ \citenamefont {Holmes}}]{QuinnGenderRoles}%
  \BibitemOpen
  \bibfield  {author} {\bibinfo {author} {\bibfnamefont {K.~N.}\ \bibnamefont
  {Quinn}}, \bibinfo {author} {\bibfnamefont {M.~M.}\ \bibnamefont {Kelley}},
  \bibinfo {author} {\bibfnamefont {K.~L.}\ \bibnamefont {McGill}}, \bibinfo
  {author} {\bibfnamefont {E.~M.}\ \bibnamefont {Smith}}, \bibinfo {author}
  {\bibfnamefont {Z.}~\bibnamefont {Whipps}},\ and\ \bibinfo {author}
  {\bibfnamefont {N.~G.}\ \bibnamefont {Holmes}},\ }\bibfield  {title}
  {\bibinfo {title} {Group roles in unstructured labs show inequitable gender
  divide},\ }\href {https://doi.org/10.1103/PhysRevPhysEducRes.16.010129}
  {\bibfield  {journal} {\bibinfo  {journal} {Phys. Rev. Phys. Educ. Res.}\
  }\textbf {\bibinfo {volume} {16}},\ \bibinfo {pages} {010129} (\bibinfo
  {year} {2020})}\BibitemShut {NoStop}%
\bibitem [{\citenamefont {Quinn}\ \emph {et~al.}(2018)\citenamefont {Quinn},
  \citenamefont {McGill}, \citenamefont {Kelley}, \citenamefont {Smith},\ and\
  \citenamefont {Holmes}}]{QuinnPERC}%
  \BibitemOpen
  \bibfield  {author} {\bibinfo {author} {\bibfnamefont {K.~N.}\ \bibnamefont
  {Quinn}}, \bibinfo {author} {\bibfnamefont {K.~L.}\ \bibnamefont {McGill}},
  \bibinfo {author} {\bibfnamefont {M.~M.}\ \bibnamefont {Kelley}}, \bibinfo
  {author} {\bibfnamefont {E.~M.}\ \bibnamefont {Smith}},\ and\ \bibinfo
  {author} {\bibfnamefont {N.~G.}\ \bibnamefont {Holmes}},\ }\bibfield  {title}
  {\bibinfo {title} {Who does what now? how physics lab instruction impacts
  student behaviors},\ }in\ \href@noop {} {\emph {\bibinfo {booktitle} {Physics
  Education Research Conference 2018}}},\ \bibinfo {series and number} {PER
  Conference}\ (\bibinfo {address} {Washington, DC},\ \bibinfo {year}
  {2018})\BibitemShut {NoStop}%
\bibitem [{\citenamefont {Dew}\ \emph {et~al.}(2022)\citenamefont {Dew},
  \citenamefont {Phillips}, \citenamefont {Karunwi}, \citenamefont {Baksh},
  \citenamefont {Stump},\ and\ \citenamefont {Holmes}}]{DewSoUnfair}%
  \BibitemOpen
  \bibfield  {author} {\bibinfo {author} {\bibfnamefont {M.}~\bibnamefont
  {Dew}}, \bibinfo {author} {\bibfnamefont {A.~M.}\ \bibnamefont {Phillips}},
  \bibinfo {author} {\bibfnamefont {S.}~\bibnamefont {Karunwi}}, \bibinfo
  {author} {\bibfnamefont {A.}~\bibnamefont {Baksh}}, \bibinfo {author}
  {\bibfnamefont {E.~M.}\ \bibnamefont {Stump}},\ and\ \bibinfo {author}
  {\bibfnamefont {N.}~\bibnamefont {Holmes}},\ }\bibfield  {title} {\bibinfo
  {title} {So unfair it’s fair: Equipment handling in remote versus in-person
  introductory physics labs},\ }in\ \href@noop {} {\emph {\bibinfo {booktitle}
  {Physics Education Research Conference, 2022}}}\ (\bibinfo {organization}
  {American Association of Physics Teachers (AAPT)},\ \bibinfo {year} {2022})\
  pp.\ \bibinfo {pages} {144--150}\BibitemShut {NoStop}%
\bibitem [{\citenamefont {Van~Domelen}(2010)}]{VanDomelenGenderEffects}%
  \BibitemOpen
  \bibfield  {author} {\bibinfo {author} {\bibfnamefont {D.}~\bibnamefont
  {Van~Domelen}},\ }\bibfield  {title} {\bibinfo {title} {Gender effects of
  computer use in a conceptual physics lab course},\ }\href@noop {} {\bibfield
  {journal} {\bibinfo  {journal} {The Physics Teacher}\ }\textbf {\bibinfo
  {volume} {48}},\ \bibinfo {pages} {534} (\bibinfo {year} {2010})}\BibitemShut
  {NoStop}%
\bibitem [{\citenamefont {Akubo}\ \emph {et~al.}(2022)\citenamefont {Akubo},
  \citenamefont {Sundstrom},\ and\ \citenamefont
  {Holmes}}]{AkuboRoleNegotiation}%
  \BibitemOpen
  \bibfield  {author} {\bibinfo {author} {\bibfnamefont {M.}~\bibnamefont
  {Akubo}}, \bibinfo {author} {\bibfnamefont {M.}~\bibnamefont {Sundstrom}},\
  and\ \bibinfo {author} {\bibfnamefont {N.~G.}\ \bibnamefont {Holmes}},\
  }\bibfield  {title} {\bibinfo {title} {Exploring diverse students'
  negotiation of lab roles through positioning},\ }in\ \href@noop {} {\emph
  {\bibinfo {booktitle} {Physics Education Research Conference 2022}}},\
  \bibinfo {series and number} {PER Conference}\ (\bibinfo {address} {Grand
  Rapids, MI},\ \bibinfo {year} {2022})\ pp.\ \bibinfo {pages}
  {30--36}\BibitemShut {NoStop}%
\bibitem [{\citenamefont {Doucette}\ \emph
  {et~al.}(2020{\natexlab{a}})\citenamefont {Doucette}, \citenamefont {Clark},\
  and\ \citenamefont {Singh}}]{DoucetteHermione}%
  \BibitemOpen
  \bibfield  {author} {\bibinfo {author} {\bibfnamefont {D.}~\bibnamefont
  {Doucette}}, \bibinfo {author} {\bibfnamefont {R.}~\bibnamefont {Clark}},\
  and\ \bibinfo {author} {\bibfnamefont {C.}~\bibnamefont {Singh}},\ }\bibfield
   {title} {\bibinfo {title} {Hermione and the secretary: How gendered task
  division in introductory physics labs can disrupt equitable learning},\
  }\href {https://doi.org/10.1088/1361-6404/ab7831} {\bibfield  {journal}
  {\bibinfo  {journal} {European Journal of Physics}\ }\textbf {\bibinfo
  {volume} {41}},\ \bibinfo {pages} {035702} (\bibinfo {year}
  {2020}{\natexlab{a}})}\BibitemShut {NoStop}%
\bibitem [{\citenamefont {Stump}\ \emph {et~al.}(2023)\citenamefont {Stump},
  \citenamefont {Dew}, \citenamefont {Jeon},\ and\ \citenamefont
  {Holmes}}]{StumpManager}%
  \BibitemOpen
  \bibfield  {author} {\bibinfo {author} {\bibfnamefont {E.~M.}\ \bibnamefont
  {Stump}}, \bibinfo {author} {\bibfnamefont {M.}~\bibnamefont {Dew}}, \bibinfo
  {author} {\bibfnamefont {S.}~\bibnamefont {Jeon}},\ and\ \bibinfo {author}
  {\bibfnamefont {N.~G.}\ \bibnamefont {Holmes}},\ }\bibfield  {title}
  {\bibinfo {title} {Taking on a manager role can support women's physics lab
  identity development},\ }\href
  {https://doi.org/10.1103/PhysRevPhysEducRes.19.010107} {\bibfield  {journal}
  {\bibinfo  {journal} {Phys. Rev. Phys. Educ. Res.}\ }\textbf {\bibinfo
  {volume} {19}},\ \bibinfo {pages} {010107} (\bibinfo {year}
  {2023})}\BibitemShut {NoStop}%
\bibitem [{\citenamefont {Doucette}\ and\ \citenamefont
  {Singh}(2021)}]{Doucette4WomenEJP}%
  \BibitemOpen
  \bibfield  {author} {\bibinfo {author} {\bibfnamefont {D.}~\bibnamefont
  {Doucette}}\ and\ \bibinfo {author} {\bibfnamefont {C.}~\bibnamefont
  {Singh}},\ }\bibfield  {title} {\bibinfo {title} {Views of female students
  who played the role of group leaders in introductory physics labs},\ }\href
  {https://doi.org/https://doi.org/10.1088/1361-6404/abd597} {\bibfield
  {journal} {\bibinfo  {journal} {European Journal of Physics}\ }\textbf
  {\bibinfo {volume} {42}},\ \bibinfo {pages} {035702} (\bibinfo {year}
  {2021})}\BibitemShut {NoStop}%
\bibitem [{\citenamefont {Holmes}\ \emph {et~al.}(2022)\citenamefont {Holmes},
  \citenamefont {Heath}, \citenamefont {Hubenig}, \citenamefont {Jeon},
  \citenamefont {Kalender}, \citenamefont {Stump},\ and\ \citenamefont
  {Sayre}}]{HolmesStudentPreference}%
  \BibitemOpen
  \bibfield  {author} {\bibinfo {author} {\bibfnamefont {N.~G.}\ \bibnamefont
  {Holmes}}, \bibinfo {author} {\bibfnamefont {G.}~\bibnamefont {Heath}},
  \bibinfo {author} {\bibfnamefont {K.}~\bibnamefont {Hubenig}}, \bibinfo
  {author} {\bibfnamefont {S.}~\bibnamefont {Jeon}}, \bibinfo {author}
  {\bibfnamefont {Z.~Y.}\ \bibnamefont {Kalender}}, \bibinfo {author}
  {\bibfnamefont {E.}~\bibnamefont {Stump}},\ and\ \bibinfo {author}
  {\bibfnamefont {E.~C.}\ \bibnamefont {Sayre}},\ }\bibfield  {title} {\bibinfo
  {title} {Evaluating the role of student preference in physics lab group
  equity},\ }\href {https://doi.org/10.1103/PhysRevPhysEducRes.18.010106}
  {\bibfield  {journal} {\bibinfo  {journal} {Phys. Rev. Phys. Educ. Res.}\
  }\textbf {\bibinfo {volume} {18}},\ \bibinfo {pages} {010106} (\bibinfo
  {year} {2022})}\BibitemShut {NoStop}%
\bibitem [{\citenamefont {Blue}\ \emph {et~al.}(2019)\citenamefont {Blue},
  \citenamefont {Traxler},\ and\ \citenamefont {Cochran}}]{BlueResourceLetter}%
  \BibitemOpen
  \bibfield  {author} {\bibinfo {author} {\bibfnamefont {J.}~\bibnamefont
  {Blue}}, \bibinfo {author} {\bibfnamefont {A.}~\bibnamefont {Traxler}},\ and\
  \bibinfo {author} {\bibfnamefont {G.}~\bibnamefont {Cochran}},\ }\bibfield
  {title} {\bibinfo {title} {Resource letter: Gp-1: Gender and physics},\
  }\href@noop {} {\bibfield  {journal} {\bibinfo  {journal} {American Journal
  of Physics}\ }\textbf {\bibinfo {volume} {87}},\ \bibinfo {pages} {616}
  (\bibinfo {year} {2019})}\BibitemShut {NoStop}%
\bibitem [{\citenamefont {Werth}\ \emph {et~al.}(2022)\citenamefont {Werth},
  \citenamefont {West},\ and\ \citenamefont {Lewandowski}}]{WerthCURE}%
  \BibitemOpen
  \bibfield  {author} {\bibinfo {author} {\bibfnamefont {A.}~\bibnamefont
  {Werth}}, \bibinfo {author} {\bibfnamefont {C.~G.}\ \bibnamefont {West}},\
  and\ \bibinfo {author} {\bibfnamefont {H.~J.}\ \bibnamefont {Lewandowski}},\
  }\bibfield  {title} {\bibinfo {title} {Impacts on student learning,
  confidence, and affect in a remote, large-enrollment, course-based
  undergraduate research experience in physics},\ }\href
  {https://doi.org/10.1103/PhysRevPhysEducRes.18.010129} {\bibfield  {journal}
  {\bibinfo  {journal} {Phys. Rev. Phys. Educ. Res.}\ }\textbf {\bibinfo
  {volume} {18}},\ \bibinfo {pages} {010129} (\bibinfo {year}
  {2022})}\BibitemShut {NoStop}%
\bibitem [{\citenamefont {Lock}\ and\ \citenamefont
  {Hazari}(2016)}]{LockDiscussing}%
  \BibitemOpen
  \bibfield  {author} {\bibinfo {author} {\bibfnamefont {R.~M.}\ \bibnamefont
  {Lock}}\ and\ \bibinfo {author} {\bibfnamefont {Z.}~\bibnamefont {Hazari}},\
  }\bibfield  {title} {\bibinfo {title} {Discussing underrepresentation as a
  means to facilitating female students' physics identity development},\ }\href
  {https://doi.org/10.1103/PhysRevPhysEducRes.12.020101} {\bibfield  {journal}
  {\bibinfo  {journal} {Phys. Rev. Phys. Educ. Res.}\ }\textbf {\bibinfo
  {volume} {12}},\ \bibinfo {pages} {020101} (\bibinfo {year}
  {2016})}\BibitemShut {NoStop}%
\bibitem [{\citenamefont {Binning}\ \emph {et~al.}(2020)\citenamefont
  {Binning}, \citenamefont {Kaufmann}, \citenamefont {McGreevy}, \citenamefont
  {Fotuhi}, \citenamefont {Chen}, \citenamefont {Marshman}, \citenamefont
  {Kalender}, \citenamefont {Limeri}, \citenamefont {Betancur},\ and\
  \citenamefont {Singh}}]{BinningChanging}%
  \BibitemOpen
  \bibfield  {author} {\bibinfo {author} {\bibfnamefont {K.~R.}\ \bibnamefont
  {Binning}}, \bibinfo {author} {\bibfnamefont {N.}~\bibnamefont {Kaufmann}},
  \bibinfo {author} {\bibfnamefont {E.~M.}\ \bibnamefont {McGreevy}}, \bibinfo
  {author} {\bibfnamefont {O.}~\bibnamefont {Fotuhi}}, \bibinfo {author}
  {\bibfnamefont {S.}~\bibnamefont {Chen}}, \bibinfo {author} {\bibfnamefont
  {E.}~\bibnamefont {Marshman}}, \bibinfo {author} {\bibfnamefont {Z.~Y.}\
  \bibnamefont {Kalender}}, \bibinfo {author} {\bibfnamefont {L.}~\bibnamefont
  {Limeri}}, \bibinfo {author} {\bibfnamefont {L.}~\bibnamefont {Betancur}},\
  and\ \bibinfo {author} {\bibfnamefont {C.}~\bibnamefont {Singh}},\ }\bibfield
   {title} {\bibinfo {title} {Changing social contexts to foster equity in
  college science courses: An ecological-belonging intervention},\ }\href@noop
  {} {\bibfield  {journal} {\bibinfo  {journal} {Psychological Science}\
  }\textbf {\bibinfo {volume} {31}},\ \bibinfo {pages} {1059} (\bibinfo {year}
  {2020})}\BibitemShut {NoStop}%
\bibitem [{\citenamefont {Doucette}\ \emph
  {et~al.}(2020{\natexlab{b}})\citenamefont {Doucette}, \citenamefont
  {D'Urso},\ and\ \citenamefont {Singh}}]{DoucetteArduinoLessons}%
  \BibitemOpen
  \bibfield  {author} {\bibinfo {author} {\bibfnamefont {D.}~\bibnamefont
  {Doucette}}, \bibinfo {author} {\bibfnamefont {B.}~\bibnamefont {D'Urso}},\
  and\ \bibinfo {author} {\bibfnamefont {C.}~\bibnamefont {Singh}},\ }\bibfield
   {title} {\bibinfo {title} {Lessons from transforming second-year honors
  physics lab},\ }\href@noop {} {\bibfield  {journal} {\bibinfo  {journal}
  {American Journal of Physics}\ }\textbf {\bibinfo {volume} {88}},\ \bibinfo
  {pages} {838} (\bibinfo {year} {2020}{\natexlab{b}})}\BibitemShut {NoStop}%
\bibitem [{\citenamefont {Gillibrand}\ \emph {et~al.}(1999)\citenamefont
  {Gillibrand}, \citenamefont {Robinson}, \citenamefont {Brawn},\ and\
  \citenamefont {Osborn}}]{GillibrandSingleSexClasses}%
  \BibitemOpen
  \bibfield  {author} {\bibinfo {author} {\bibfnamefont {E.}~\bibnamefont
  {Gillibrand}}, \bibinfo {author} {\bibfnamefont {P.}~\bibnamefont
  {Robinson}}, \bibinfo {author} {\bibfnamefont {R.}~\bibnamefont {Brawn}},\
  and\ \bibinfo {author} {\bibfnamefont {A.}~\bibnamefont {Osborn}},\
  }\bibfield  {title} {\bibinfo {title} {Girls' participation in physics in
  single sex classes in mixed schools in relation to confidence and
  achievement},\ }\href@noop {} {\bibfield  {journal} {\bibinfo  {journal}
  {International Journal of Science Education}\ }\textbf {\bibinfo {volume}
  {21}},\ \bibinfo {pages} {349} (\bibinfo {year} {1999})}\BibitemShut
  {NoStop}%
\bibitem [{\citenamefont {Heller}\ and\ \citenamefont
  {Hollabaugh}(1992)}]{HellerGroupwork}%
  \BibitemOpen
  \bibfield  {author} {\bibinfo {author} {\bibfnamefont {P.}~\bibnamefont
  {Heller}}\ and\ \bibinfo {author} {\bibfnamefont {M.}~\bibnamefont
  {Hollabaugh}},\ }\bibfield  {title} {\bibinfo {title} {Teaching problem
  solving through cooperative grouping. {P}art 2: Designing problems and
  structuring groups},\ }\href@noop {} {\bibfield  {journal} {\bibinfo
  {journal} {Am. J. Phys.}\ }\textbf {\bibinfo {volume} {60}},\ \bibinfo
  {pages} {637} (\bibinfo {year} {1992})}\BibitemShut {NoStop}%
\bibitem [{\citenamefont {McCullough}(2007)}]{McCulloughGenderClassroom}%
  \BibitemOpen
  \bibfield  {author} {\bibinfo {author} {\bibfnamefont {L.}~\bibnamefont
  {McCullough}},\ }\bibfield  {title} {\bibinfo {title} {Gender in the physics
  classroom},\ }\href@noop {} {\bibfield  {journal} {\bibinfo  {journal} {The
  Physics Teacher}\ }\textbf {\bibinfo {volume} {45}},\ \bibinfo {pages} {316}
  (\bibinfo {year} {2007})}\BibitemShut {NoStop}%
\bibitem [{\citenamefont {Brannen}\ \emph {et~al.}(2021)\citenamefont
  {Brannen}, \citenamefont {Beauchamp}, \citenamefont {Cartwright},
  \citenamefont {Liddle}, \citenamefont {Tishinsky}, \citenamefont {Newton},\
  and\ \citenamefont {Monk}}]{BrannenGroupworkContract}%
  \BibitemOpen
  \bibfield  {author} {\bibinfo {author} {\bibfnamefont {S.~F.}\ \bibnamefont
  {Brannen}}, \bibinfo {author} {\bibfnamefont {D.}~\bibnamefont {Beauchamp}},
  \bibinfo {author} {\bibfnamefont {N.~M.}\ \bibnamefont {Cartwright}},
  \bibinfo {author} {\bibfnamefont {D.~M.}\ \bibnamefont {Liddle}}, \bibinfo
  {author} {\bibfnamefont {J.~M.}\ \bibnamefont {Tishinsky}}, \bibinfo {author}
  {\bibfnamefont {G.}~\bibnamefont {Newton}},\ and\ \bibinfo {author}
  {\bibfnamefont {J.~M.}\ \bibnamefont {Monk}},\ }\bibfield  {title} {\bibinfo
  {title} {Effectiveness of group work contracts to facilitate collaborative
  group learning and reduce anxiety in traditional face-to-face lecture and
  online distance education course formats.},\ }\href@noop {} {\bibfield
  {journal} {\bibinfo  {journal} {International Journal for the Scholarship of
  Teaching and Learning}\ }\textbf {\bibinfo {volume} {15}},\ \bibinfo {pages}
  {5} (\bibinfo {year} {2021})}\BibitemShut {NoStop}%
\bibitem [{\citenamefont {Lewis}\ \emph {et~al.}(2019)\citenamefont {Lewis},
  \citenamefont {Sekaquaptewa},\ and\ \citenamefont
  {Meadows}}]{LewisIntervention}%
  \BibitemOpen
  \bibfield  {author} {\bibinfo {author} {\bibfnamefont {N.~A.}\ \bibnamefont
  {Lewis}}, \bibinfo {author} {\bibfnamefont {D.}~\bibnamefont
  {Sekaquaptewa}},\ and\ \bibinfo {author} {\bibfnamefont {L.~A.}\ \bibnamefont
  {Meadows}},\ }\bibfield  {title} {\bibinfo {title} {Modeling gender
  counter-stereotypic group behavior: A brief video intervention reduces
  participation gender gaps on {STEM} teams},\ }\href@noop {} {\bibfield
  {journal} {\bibinfo  {journal} {Social Psychology of Education}\ }\textbf
  {\bibinfo {volume} {22}},\ \bibinfo {pages} {557} (\bibinfo {year}
  {2019})}\BibitemShut {NoStop}%
\bibitem [{\citenamefont {{AAPT Committee on Laboratories}}(2015)}]{AAPTlab}%
  \BibitemOpen
  \bibfield  {author} {\bibinfo {author} {\bibnamefont {{AAPT Committee on
  Laboratories}}},\ }\href@noop {} {\emph {\bibinfo {title} {AAPT
  Recommendations for the Undergraduate Physics Laboratory Curriculum}}}\
  (\bibinfo  {publisher} {AAPT},\ \bibinfo {year} {2015})\BibitemShut {NoStop}%
\bibitem [{\citenamefont {Callan}\ \emph {et~al.}(2017)\citenamefont {Callan},
  \citenamefont {Wilcox},\ and\ \citenamefont {Adams}}]{CallanGroups}%
  \BibitemOpen
  \bibfield  {author} {\bibinfo {author} {\bibfnamefont {K.~E.}\ \bibnamefont
  {Callan}}, \bibinfo {author} {\bibfnamefont {B.}~\bibnamefont {Wilcox}},\
  and\ \bibinfo {author} {\bibfnamefont {W.}~\bibnamefont {Adams}},\ }\bibfield
   {title} {\bibinfo {title} {Testing group composition within a studio
  learning environment},\ }in\ \href@noop {} {\emph {\bibinfo {booktitle}
  {Physics Education Research Conference 2017}}},\ \bibinfo {series and number}
  {PER Conference}\ (\bibinfo {address} {Cincinnati, OH},\ \bibinfo {year}
  {2017})\ pp.\ \bibinfo {pages} {72--75}\BibitemShut {NoStop}%
\bibitem [{\citenamefont {Heller}\ and\ \citenamefont
  {Heller}(2001)}]{HellerCooperative}%
  \BibitemOpen
  \bibfield  {author} {\bibinfo {author} {\bibfnamefont {P.}~\bibnamefont
  {Heller}}\ and\ \bibinfo {author} {\bibfnamefont {K.}~\bibnamefont
  {Heller}},\ }\href@noop {} {\emph {\bibinfo {title} {Cooperative Group
  Problem Solving in Physics}}}\ (\bibinfo  {publisher} {Brooks/Cole Publishing
  Company New York},\ \bibinfo {year} {2001})\BibitemShut {NoStop}%
\bibitem [{\citenamefont {Doucette}\ \emph
  {et~al.}(2020{\natexlab{c}})\citenamefont {Doucette}, \citenamefont {Clark},\
  and\ \citenamefont {Singh}}]{DoucetteGoodLabPartnerPERC}%
  \BibitemOpen
  \bibfield  {author} {\bibinfo {author} {\bibfnamefont {D.}~\bibnamefont
  {Doucette}}, \bibinfo {author} {\bibfnamefont {R.}~\bibnamefont {Clark}},\
  and\ \bibinfo {author} {\bibfnamefont {C.}~\bibnamefont {Singh}},\ }\bibfield
   {title} {\bibinfo {title} {What makes a good physics lab partner?},\ }in\
  \href@noop {} {\emph {\bibinfo {booktitle} {Physics Education Research
  Conference 2020}}},\ \bibinfo {series and number} {PER Conference}\ (\bibinfo
  {address} {Virtual Conference},\ \bibinfo {year} {2020})\ pp.\ \bibinfo
  {pages} {124--130}\BibitemShut {NoStop}%
\bibitem [{\citenamefont {Doucette}\ and\ \citenamefont
  {Singh}(2022)}]{DoucetteShareIt}%
  \BibitemOpen
  \bibfield  {author} {\bibinfo {author} {\bibfnamefont {D.}~\bibnamefont
  {Doucette}}\ and\ \bibinfo {author} {\bibfnamefont {C.}~\bibnamefont
  {Singh}},\ }\bibfield  {title} {\bibinfo {title} {Share it, don’t split it:
  Can equitable group work improve student outcomes?},\ }\href@noop {}
  {\bibfield  {journal} {\bibinfo  {journal} {The Physics Teacher}\ }\textbf
  {\bibinfo {volume} {60}},\ \bibinfo {pages} {166} (\bibinfo {year}
  {2022})}\BibitemShut {NoStop}%
\bibitem [{\citenamefont {Johnson}(2020)}]{JohnsonIntersectionalFramework}%
  \BibitemOpen
  \bibfield  {author} {\bibinfo {author} {\bibfnamefont {A.}~\bibnamefont
  {Johnson}},\ }\bibfield  {title} {\bibinfo {title} {An intersectional physics
  identity framework for studying physics settings},\ }in\ \href@noop {} {\emph
  {\bibinfo {booktitle} {Physics education and gender: identity as an analytic
  lens for research}}},\ \bibinfo {editor} {edited by\ \bibinfo {editor}
  {\bibfnamefont {A.~J.}\ \bibnamefont {Gonsalves}}\ and\ \bibinfo {editor}
  {\bibfnamefont {A.~T.}\ \bibnamefont {Danielsson}}}\ (\bibinfo  {publisher}
  {Springer},\ \bibinfo {year} {2020})\ pp.\ \bibinfo {pages}
  {53--80}\BibitemShut {NoStop}%
\bibitem [{\citenamefont {Marshman}\ \emph {et~al.}(2018)\citenamefont
  {Marshman}, \citenamefont {Kalender}, \citenamefont {Nokes-Malach},
  \citenamefont {Schunn},\ and\ \citenamefont {Singh}}]{MarshmanSE}%
  \BibitemOpen
  \bibfield  {author} {\bibinfo {author} {\bibfnamefont {E.~M.}\ \bibnamefont
  {Marshman}}, \bibinfo {author} {\bibfnamefont {Z.~Y.}\ \bibnamefont
  {Kalender}}, \bibinfo {author} {\bibfnamefont {T.}~\bibnamefont
  {Nokes-Malach}}, \bibinfo {author} {\bibfnamefont {C.}~\bibnamefont
  {Schunn}},\ and\ \bibinfo {author} {\bibfnamefont {C.}~\bibnamefont
  {Singh}},\ }\bibfield  {title} {\bibinfo {title} {{Female students with A's
  have similar physics self-efficacy as male students with C's in introductory
  courses: A cause for alarm?}},\ }\href
  {https://doi.org/10.1103/PhysRevPhysEducRes.14.020123} {\bibfield  {journal}
  {\bibinfo  {journal} {Phys. Rev. Phys. Educ. Res.}\ }\textbf {\bibinfo
  {volume} {14}},\ \bibinfo {pages} {020123} (\bibinfo {year}
  {2018})}\BibitemShut {NoStop}%
\bibitem [{\citenamefont {Kalender}\ \emph {et~al.}(2019)\citenamefont
  {Kalender}, \citenamefont {Marshman}, \citenamefont {Schunn}, \citenamefont
  {Nokes-Malach},\ and\ \citenamefont {Singh}}]{KalenderIdentity}%
  \BibitemOpen
  \bibfield  {author} {\bibinfo {author} {\bibfnamefont {Z.~Y.}\ \bibnamefont
  {Kalender}}, \bibinfo {author} {\bibfnamefont {E.}~\bibnamefont {Marshman}},
  \bibinfo {author} {\bibfnamefont {C.~D.}\ \bibnamefont {Schunn}}, \bibinfo
  {author} {\bibfnamefont {T.~J.}\ \bibnamefont {Nokes-Malach}},\ and\ \bibinfo
  {author} {\bibfnamefont {C.}~\bibnamefont {Singh}},\ }\bibfield  {title}
  {\bibinfo {title} {Why female science, technology, engineering, and
  mathematics majors do not identify with physics: They do not think others see
  them that way},\ }\href
  {https://doi.org/10.1103/PhysRevPhysEducRes.15.020148} {\bibfield  {journal}
  {\bibinfo  {journal} {Phys. Rev. Phys. Educ. Res.}\ }\textbf {\bibinfo
  {volume} {15}},\ \bibinfo {pages} {020148} (\bibinfo {year}
  {2019})}\BibitemShut {NoStop}%
\bibitem [{\citenamefont {Hazari}\ \emph {et~al.}(2020)\citenamefont {Hazari},
  \citenamefont {Chari}, \citenamefont {Potvin},\ and\ \citenamefont
  {Brewe}}]{HazariIdentity}%
  \BibitemOpen
  \bibfield  {author} {\bibinfo {author} {\bibfnamefont {Z.}~\bibnamefont
  {Hazari}}, \bibinfo {author} {\bibfnamefont {D.}~\bibnamefont {Chari}},
  \bibinfo {author} {\bibfnamefont {G.}~\bibnamefont {Potvin}},\ and\ \bibinfo
  {author} {\bibfnamefont {E.}~\bibnamefont {Brewe}},\ }\bibfield  {title}
  {\bibinfo {title} {The context dependence of physics identity: Examining the
  role of performance/competence, recognition, interest, and sense of belonging
  for lower and upper female physics undergraduates},\ }\href@noop {}
  {\bibfield  {journal} {\bibinfo  {journal} {Journal of Research in Science
  Teaching}\ }\textbf {\bibinfo {volume} {57}},\ \bibinfo {pages} {1583}
  (\bibinfo {year} {2020})}\BibitemShut {NoStop}%
\bibitem [{\citenamefont {Gosling}(2004)}]{GoslingLabChoice}%
  \BibitemOpen
  \bibfield  {author} {\bibinfo {author} {\bibfnamefont {C.}~\bibnamefont
  {Gosling}},\ }\bibfield  {title} {\bibinfo {title} {Challenges facing high
  school physics students: An annotated synopsis of peer-reviewed literature
  addressing curriculum relevance and gender},\ }\href@noop {} {\bibfield
  {journal} {\bibinfo  {journal} {Journal of Physics Teacher Education Online}\
  }\textbf {\bibinfo {volume} {2}},\ \bibinfo {pages} {1} (\bibinfo {year}
  {2004})}\BibitemShut {NoStop}%
\bibitem [{\citenamefont {Ambrosia}(1940)}]{AmbrosiaTeachingPhysicsToWomen}%
  \BibitemOpen
  \bibfield  {author} {\bibinfo {author} {\bibfnamefont {S.~M.}\ \bibnamefont
  {Ambrosia}},\ }\bibfield  {title} {\bibinfo {title} {Teaching physics to
  women},\ }\href {https://doi.org/10.1119/1.1991588} {\bibfield  {journal}
  {\bibinfo  {journal} {American Journal of Physics}\ }\textbf {\bibinfo
  {volume} {8}},\ \bibinfo {pages} {289} (\bibinfo {year} {1940})}\BibitemShut
  {NoStop}%
\bibitem [{\citenamefont {Turpen}\ \emph {et~al.}(2018)\citenamefont {Turpen},
  \citenamefont {Gupta}, \citenamefont {Radoff}, \citenamefont {Elby},
  \citenamefont {Sabo},\ and\ \citenamefont {Quan}}]{TurpenGroupInequity}%
  \BibitemOpen
  \bibfield  {author} {\bibinfo {author} {\bibfnamefont {C.}~\bibnamefont
  {Turpen}}, \bibinfo {author} {\bibfnamefont {A.}~\bibnamefont {Gupta}},
  \bibinfo {author} {\bibfnamefont {J.}~\bibnamefont {Radoff}}, \bibinfo
  {author} {\bibfnamefont {A.}~\bibnamefont {Elby}}, \bibinfo {author}
  {\bibfnamefont {H.}~\bibnamefont {Sabo}},\ and\ \bibinfo {author}
  {\bibfnamefont {G.}~\bibnamefont {Quan}},\ }\bibfield  {title} {\bibinfo
  {title} {Successes and challenges in supporting undergraduate peer educators
  to notice and respond to equity considerations within design teams},\ }in\
  \href@noop {} {\emph {\bibinfo {booktitle} {2018 ASEE Annual Conference and
  Exposition}}}\ (\bibinfo {address} {Salt Lake City, UT},\ \bibinfo {year}
  {2018})\BibitemShut {NoStop}%
\bibitem [{\citenamefont {Alicea-Mu\~noz}\ \emph {et~al.}(2021)\citenamefont
  {Alicea-Mu\~noz}, \citenamefont {Subi\~no Sullivan},\ and\ \citenamefont
  {Schatz}}]{MunozTAs}%
  \BibitemOpen
  \bibfield  {author} {\bibinfo {author} {\bibfnamefont {E.}~\bibnamefont
  {Alicea-Mu\~noz}}, \bibinfo {author} {\bibfnamefont {C.}~\bibnamefont
  {Subi\~no Sullivan}},\ and\ \bibinfo {author} {\bibfnamefont {M.~F.}\
  \bibnamefont {Schatz}},\ }\bibfield  {title} {\bibinfo {title} {Transforming
  the preparation of physics graduate teaching assistants: Curriculum
  development},\ }\href {https://doi.org/10.1103/PhysRevPhysEducRes.17.020125}
  {\bibfield  {journal} {\bibinfo  {journal} {Phys. Rev. Phys. Educ. Res.}\
  }\textbf {\bibinfo {volume} {17}},\ \bibinfo {pages} {020125} (\bibinfo
  {year} {2021})}\BibitemShut {NoStop}%
\bibitem [{\citenamefont {Wan}\ \emph {et~al.}(2021)\citenamefont {Wan},
  \citenamefont {Doty}, \citenamefont {Geraets}, \citenamefont {Nix},
  \citenamefont {Saitta},\ and\ \citenamefont {Chini}}]{WanTAs}%
  \BibitemOpen
  \bibfield  {author} {\bibinfo {author} {\bibfnamefont {T.}~\bibnamefont
  {Wan}}, \bibinfo {author} {\bibfnamefont {C.~M.}\ \bibnamefont {Doty}},
  \bibinfo {author} {\bibfnamefont {A.~A.}\ \bibnamefont {Geraets}}, \bibinfo
  {author} {\bibfnamefont {C.~A.}\ \bibnamefont {Nix}}, \bibinfo {author}
  {\bibfnamefont {E.~K.~H.}\ \bibnamefont {Saitta}},\ and\ \bibinfo {author}
  {\bibfnamefont {J.~J.}\ \bibnamefont {Chini}},\ }\bibfield  {title} {\bibinfo
  {title} {Evaluating the impact of a classroom simulator training on graduate
  teaching assistants' instructional practices and undergraduate student
  learning},\ }\href {https://doi.org/10.1103/PhysRevPhysEducRes.17.010146}
  {\bibfield  {journal} {\bibinfo  {journal} {Phys. Rev. Phys. Educ. Res.}\
  }\textbf {\bibinfo {volume} {17}},\ \bibinfo {pages} {010146} (\bibinfo
  {year} {2021})}\BibitemShut {NoStop}%
\bibitem [{\citenamefont {Doucette}\ \emph
  {et~al.}(2020{\natexlab{d}})\citenamefont {Doucette}, \citenamefont {Clark},\
  and\ \citenamefont {Singh}}]{DoucettePDforLabTAs}%
  \BibitemOpen
  \bibfield  {author} {\bibinfo {author} {\bibfnamefont {D.}~\bibnamefont
  {Doucette}}, \bibinfo {author} {\bibfnamefont {R.}~\bibnamefont {Clark}},\
  and\ \bibinfo {author} {\bibfnamefont {C.}~\bibnamefont {Singh}},\ }\bibfield
   {title} {\bibinfo {title} {Professional development combining cognitive
  apprenticeship and expectancy-value theories improves lab teaching
  assistants' instructional views and practices},\ }\href
  {https://doi.org/10.1103/PhysRevPhysEducRes.16.020102} {\bibfield  {journal}
  {\bibinfo  {journal} {Phys. Rev. Phys. Educ. Res.}\ }\textbf {\bibinfo
  {volume} {16}},\ \bibinfo {pages} {020102} (\bibinfo {year}
  {2020}{\natexlab{d}})}\BibitemShut {NoStop}%
\bibitem [{\citenamefont {Freeman}\ \emph {et~al.}(2014)\citenamefont
  {Freeman}, \citenamefont {Eddy}, \citenamefont {McDonough}, \citenamefont
  {Smith}, \citenamefont {Okoroafor}, \citenamefont {Jordt},\ and\
  \citenamefont {Wenderoth}}]{FreemanActiveLearning}%
  \BibitemOpen
  \bibfield  {author} {\bibinfo {author} {\bibfnamefont {S.}~\bibnamefont
  {Freeman}}, \bibinfo {author} {\bibfnamefont {S.~L.}\ \bibnamefont {Eddy}},
  \bibinfo {author} {\bibfnamefont {M.}~\bibnamefont {McDonough}}, \bibinfo
  {author} {\bibfnamefont {M.~K.}\ \bibnamefont {Smith}}, \bibinfo {author}
  {\bibfnamefont {N.}~\bibnamefont {Okoroafor}}, \bibinfo {author}
  {\bibfnamefont {H.}~\bibnamefont {Jordt}},\ and\ \bibinfo {author}
  {\bibfnamefont {M.~P.}\ \bibnamefont {Wenderoth}},\ }\bibfield  {title}
  {\bibinfo {title} {Active learning increases student performance in science,
  engineering, and mathematics},\ }\href@noop {} {\bibfield  {journal}
  {\bibinfo  {journal} {Proceedings of the National Academy of Sciences}\
  }\textbf {\bibinfo {volume} {111}},\ \bibinfo {pages} {8410} (\bibinfo {year}
  {2014})}\BibitemShut {NoStop}%
\bibitem [{\citenamefont {Espinosa}\ \emph {et~al.}(2019)\citenamefont
  {Espinosa}, \citenamefont {Miller}, \citenamefont {Araujo},\ and\
  \citenamefont {Mazur}}]{EspinosaGenderGap}%
  \BibitemOpen
  \bibfield  {author} {\bibinfo {author} {\bibfnamefont {T.}~\bibnamefont
  {Espinosa}}, \bibinfo {author} {\bibfnamefont {K.}~\bibnamefont {Miller}},
  \bibinfo {author} {\bibfnamefont {I.}~\bibnamefont {Araujo}},\ and\ \bibinfo
  {author} {\bibfnamefont {E.}~\bibnamefont {Mazur}},\ }\bibfield  {title}
  {\bibinfo {title} {Reducing the gender gap in students' physics self-efficacy
  in a team- and project-based introductory physics class},\ }\href
  {https://doi.org/10.1103/PhysRevPhysEducRes.15.010132} {\bibfield  {journal}
  {\bibinfo  {journal} {Phys. Rev. Phys. Educ. Res.}\ }\textbf {\bibinfo
  {volume} {15}},\ \bibinfo {pages} {010132} (\bibinfo {year}
  {2019})}\BibitemShut {NoStop}%
\bibitem [{\citenamefont {Lorenzo}\ \emph {et~al.}(2006)\citenamefont
  {Lorenzo}, \citenamefont {Crouch},\ and\ \citenamefont
  {Mazur}}]{LorenzoCrouchMazur}%
  \BibitemOpen
  \bibfield  {author} {\bibinfo {author} {\bibfnamefont {M.}~\bibnamefont
  {Lorenzo}}, \bibinfo {author} {\bibfnamefont {C.~H.}\ \bibnamefont
  {Crouch}},\ and\ \bibinfo {author} {\bibfnamefont {E.}~\bibnamefont
  {Mazur}},\ }\bibfield  {title} {\bibinfo {title} {Reducing the gender gap in
  the physics classroom},\ }\href {https://doi.org/10.1119/1.2162549}
  {\bibfield  {journal} {\bibinfo  {journal} {American Journal of Physics}\
  }\textbf {\bibinfo {volume} {74}},\ \bibinfo {pages} {118} (\bibinfo {year}
  {2006})}\BibitemShut {NoStop}%
\bibitem [{\citenamefont {Good}\ \emph {et~al.}(2019)\citenamefont {Good},
  \citenamefont {Maries},\ and\ \citenamefont
  {Singh}}]{GoodGenderProblemSolving}%
  \BibitemOpen
  \bibfield  {author} {\bibinfo {author} {\bibfnamefont {M.}~\bibnamefont
  {Good}}, \bibinfo {author} {\bibfnamefont {A.}~\bibnamefont {Maries}},\ and\
  \bibinfo {author} {\bibfnamefont {C.}~\bibnamefont {Singh}},\ }\bibfield
  {title} {\bibinfo {title} {Impact of traditional or evidence-based
  active-engagement instruction on introductory female and male students'
  attitudes and approaches to physics problem solving},\ }\href
  {https://doi.org/10.1103/PhysRevPhysEducRes.15.020129} {\bibfield  {journal}
  {\bibinfo  {journal} {Phys. Rev. Phys. Educ. Res.}\ }\textbf {\bibinfo
  {volume} {15}},\ \bibinfo {pages} {020129} (\bibinfo {year}
  {2019})}\BibitemShut {NoStop}%
\bibitem [{\citenamefont {Karim}\ \emph {et~al.}(2018)\citenamefont {Karim},
  \citenamefont {Maries},\ and\ \citenamefont {Singh}}]{KarimEBAEgender}%
  \BibitemOpen
  \bibfield  {author} {\bibinfo {author} {\bibfnamefont {N.~I.}\ \bibnamefont
  {Karim}}, \bibinfo {author} {\bibfnamefont {A.}~\bibnamefont {Maries}},\ and\
  \bibinfo {author} {\bibfnamefont {C.}~\bibnamefont {Singh}},\ }\bibfield
  {title} {\bibinfo {title} {Do evidence-based active-engagement courses reduce
  the gender gap in introductory physics?},\ }\href@noop {} {\bibfield
  {journal} {\bibinfo  {journal} {Eur. J. Phys.}\ }\textbf {\bibinfo {volume}
  {39}},\ \bibinfo {pages} {025701} (\bibinfo {year} {2018})}\BibitemShut
  {NoStop}%
\bibitem [{\citenamefont {Maries}\ \emph {et~al.}(2020)\citenamefont {Maries},
  \citenamefont {Karim},\ and\ \citenamefont {Singh}}]{MariesStereotypeThreat}%
  \BibitemOpen
  \bibfield  {author} {\bibinfo {author} {\bibfnamefont {A.}~\bibnamefont
  {Maries}}, \bibinfo {author} {\bibfnamefont {N.~I.}\ \bibnamefont {Karim}},\
  and\ \bibinfo {author} {\bibfnamefont {C.}~\bibnamefont {Singh}},\ }\bibfield
   {title} {\bibinfo {title} {Active learning in an inequitable learning
  environment can increase the gender performance gap: The negative impact of
  stereotype threat},\ }\href@noop {} {\bibfield  {journal} {\bibinfo
  {journal} {The Physics Teacher}\ }\textbf {\bibinfo {volume} {58}},\ \bibinfo
  {pages} {430} (\bibinfo {year} {2020})}\BibitemShut {NoStop}%
\bibitem [{\citenamefont {Etkina}\ \emph {et~al.}(1999)\citenamefont {Etkina},
  \citenamefont {Gibbons}, \citenamefont {Holton},\ and\ \citenamefont
  {Horton}}]{EtkinaLessonsLearned}%
  \BibitemOpen
  \bibfield  {author} {\bibinfo {author} {\bibfnamefont {E.}~\bibnamefont
  {Etkina}}, \bibinfo {author} {\bibfnamefont {K.}~\bibnamefont {Gibbons}},
  \bibinfo {author} {\bibfnamefont {B.}~\bibnamefont {Holton}},\ and\ \bibinfo
  {author} {\bibfnamefont {G.}~\bibnamefont {Horton}},\ }\bibfield  {title}
  {\bibinfo {title} {Lessons learned: A case study of an integrated way of
  teaching introductory physics to at-risk students at rutgers university},\
  }\href@noop {} {\bibfield  {journal} {\bibinfo  {journal} {American Journal
  of Physics}\ }\textbf {\bibinfo {volume} {67}},\ \bibinfo {pages} {810}
  (\bibinfo {year} {1999})}\BibitemShut {NoStop}%
\bibitem [{\citenamefont {Dounas-Frazer}\ \emph {et~al.}(2022)\citenamefont
  {Dounas-Frazer}, \citenamefont {Gillen}, \citenamefont {Herne}, \citenamefont
  {Howard}, \citenamefont {Lindell}, \citenamefont {McGrew}, \citenamefont
  {Mumford}, \citenamefont {Nguyen}, \citenamefont {Osadchuk}, \citenamefont
  {Crane} \emph {et~al.}}]{DounasFrazerAccessibleLabs}%
  \BibitemOpen
  \bibfield  {author} {\bibinfo {author} {\bibfnamefont {D.~R.}\ \bibnamefont
  {Dounas-Frazer}}, \bibinfo {author} {\bibfnamefont {D.}~\bibnamefont
  {Gillen}}, \bibinfo {author} {\bibfnamefont {C.~M.}\ \bibnamefont {Herne}},
  \bibinfo {author} {\bibfnamefont {E.}~\bibnamefont {Howard}}, \bibinfo
  {author} {\bibfnamefont {R.~S.}\ \bibnamefont {Lindell}}, \bibinfo {author}
  {\bibfnamefont {G.~I.}\ \bibnamefont {McGrew}}, \bibinfo {author}
  {\bibfnamefont {J.~R.}\ \bibnamefont {Mumford}}, \bibinfo {author}
  {\bibfnamefont {N.~H.}\ \bibnamefont {Nguyen}}, \bibinfo {author}
  {\bibfnamefont {L.}~\bibnamefont {Osadchuk}}, \bibinfo {author}
  {\bibfnamefont {J.~P.}\ \bibnamefont {Crane}}, \emph {et~al.},\ }\href@noop
  {} {\emph {\bibinfo {title} {Increase Investment in Accessible Physics Labs:
  A Call to Action for the Physics Education Community}}},\ \bibinfo {type}
  {Tech. Rep.}\ (\bibinfo  {institution} {American Association of Physics
  Teachers},\ \bibinfo {year} {2022})\BibitemShut {NoStop}%
\end{thebibliography}%

\end{document}